# Global Survey of Technologies and Industrial Applications of Grid Forming Energy Storage Systems

Heng Wu, *Senior Member*, *IEEE*, Changjiang Zhan, Jiacheng Li, Xiaoyao Zhou, and Xiongfei Wang, *Fellow*, *IEEE*

***Abstract*—Grid-forming (GFM) energy storage system (ESS) is a key enabler for stabilizing future power systems with high penetration of converter-based resources (CBRs). To get a better overview of the state-of-the-art and challenges for implementing and deploying GFM-ESS, a global survey has been initiated by Cigre Working Group B4.101 - industrial implementation and application of grid forming energy storage systems. Feedback was collected from universities, transmission system operators (TSOs), power plant developers, original equipment manufacturers (OEMs), research institutes, as well as consultants. It is interesting to note that while many common understandings have been established in practice, certain gaps persist among different stakeholders. This article intends to bridge this gap by presenting a summary of the survey, including the questionnaire, responses from various stakeholders, and in-depth analysis of the survey results. The key challenges faced by different stakeholders in deploying GFM-ESS are identified, shedding light on future research in this direction.**

## I. Introduction

The grid-forming (GFM) technology is emerging as a promising approach for massive integration of converter-based resources (CBRs) into electrical grids [1]-[2]. Being controlled as a voltage source behind an impedance, GFM-CBRs can provide adequate services to enhance the reliability and resilience of the power network, and they also feature higher stability robustness against grid strength variations compared to conventional grid-following (GFL) CBRs [3]-[4]. Among other applications, implementing GFM control in the energy storage system (ESS) is simpler and more straightforward, as most GFM controls are designed with the assumptions of stable DC-link voltage and sufficient energy buffer [5]-[6]. Those assumptions are typically met when the primary power source of the converter is an energy storage unit, e.g., a battery. Consequently, a substantial number of real-world projects utilizing GFM control are based on ESSs, making GFM-ESS a frontrunner in the practical deployment of this technology [7].

Over past years, numerous efforts have been devoted, from both academia and industry, to advance GFM technology in different applications [8]-[12]. However, certain gaps persist among different stakeholders. An over-simplified CBR-grid system with a few or even single GFM-ESS is often used by academia for dynamic analysis and controller design. Such a simplified system may not be capable of reflecting all practical challenges in complex electrical systems, where thousands of GFM-CBRs can be configured in meshed and radial network, and interact with GFL-CBRs and synchronous generators. On the other hand, practicing engineers may not be aware of the latest progresses in academia in terms of modeling, stability analysis, and control of GFM-ESS.

To bridge the gaps, Cigre B4.101- industrial implementation and application of grid forming energy storage systems [13], has initiated a global survey of technologies and industrial applications of GFM-ESS, aiming to obtain insights into the latest state-of-the-art and challenges on this topic. The questionnaire has 16 questions in total, and 105 responses were received from a diverse range of participants, including universities, transmission system operators (TSOs), power plant developers, original equipment manufacturers (OEMs), research institutes, as well as consultants, as shown in Table I and II. More detailed statistics of the survey participants are given in Table A-I to Table A-VI in the appendix.

TABLE I
RESPONDENT SECTOR STATISTICS

| Sector | Number |
|---|---|
| **Academia** | 18 |
| **Consultant** | 8 |
| **Developer** | 7 |
| **OEM** | 34 |
| **Research Institute** | 13 |
| **TSO/DSO** | 25 |
| **Total** | 105 |

TABLE II
GEOGRAPHIC LOCATION OF RESPONDENTS.

| Continent | Number |
|---|---|
| **Africa** | 2 |
| **Asia** | 30 |
| **Europe** | 51 |
| **North America** | 11 |
| **Oceania** | 9 |
| **South America** | 2 |

This article intends to share the key results of this global survey. It summarizes the common understandings, as well as the challenges faced by different stakeholders in the development, deployment, and operation of GFM-ESS. By presenting a structured analysis of the questionnaire responses, this paper aims to provide a clear picture of the current industrial landscape and identify critical areas for future research and collaborative efforts, thereby accelerating the adoption of GFM-ESS technology.

## II. Questionnaire and Responses

The questionnaire is designed with 16 technical questions, each question and the corresponding response is listed as follows:

1. **According to you, which statement best fits the definition of the GFM *functionality*? (single choice)**
   A. GFM refers to a specific set of features that characterize the transient response of a grid-connected device (e.g. being able to provide true inertia, short-term system strength/voltage amplitude stiffness, system damping, stable operation after loss of last synchronous machine, etc).
   B. GFM refers to a specific control that can be implemented in converter-based resources (CBRs) to mimic the behavior of synchronous machines (e.g. virtual synchronous machine control).
   C. GFM refers to the removal of the classical phase-locked loop (PLL) synchronization in CBR control (therefore when a PLL is used, independently of the purpose, the device cannot be called GFM).
   D. Others (Please specify if you don't agree with all above options):

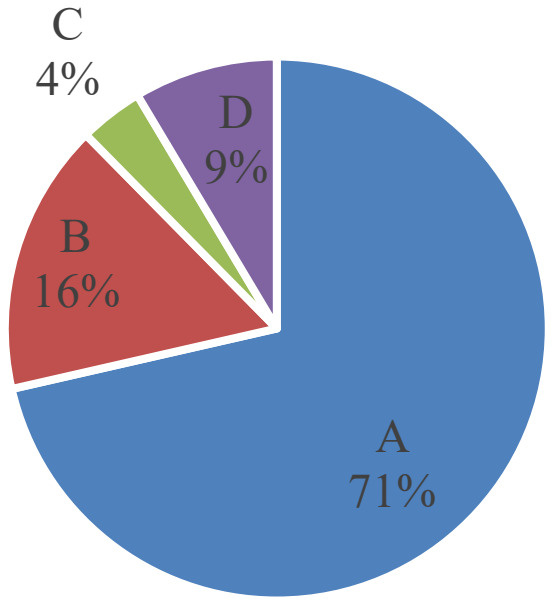

Fig. 1. Response to Q1

2. **Are the following features included in the "core" GFM *functionality*? (Multiple choice)**
   A. Capability that requires sufficient energy buffer, such as black start capability, islanding operation that sustains load for minutes or hours, etc.
   B. Short-term overload capability to contribute to fault level.
   C. None of the above are "core" GFM functionalities, but could be "additional/non-mandatory" GFM functionalities.
   D. Others (Please specify if you don't agree with all above options):

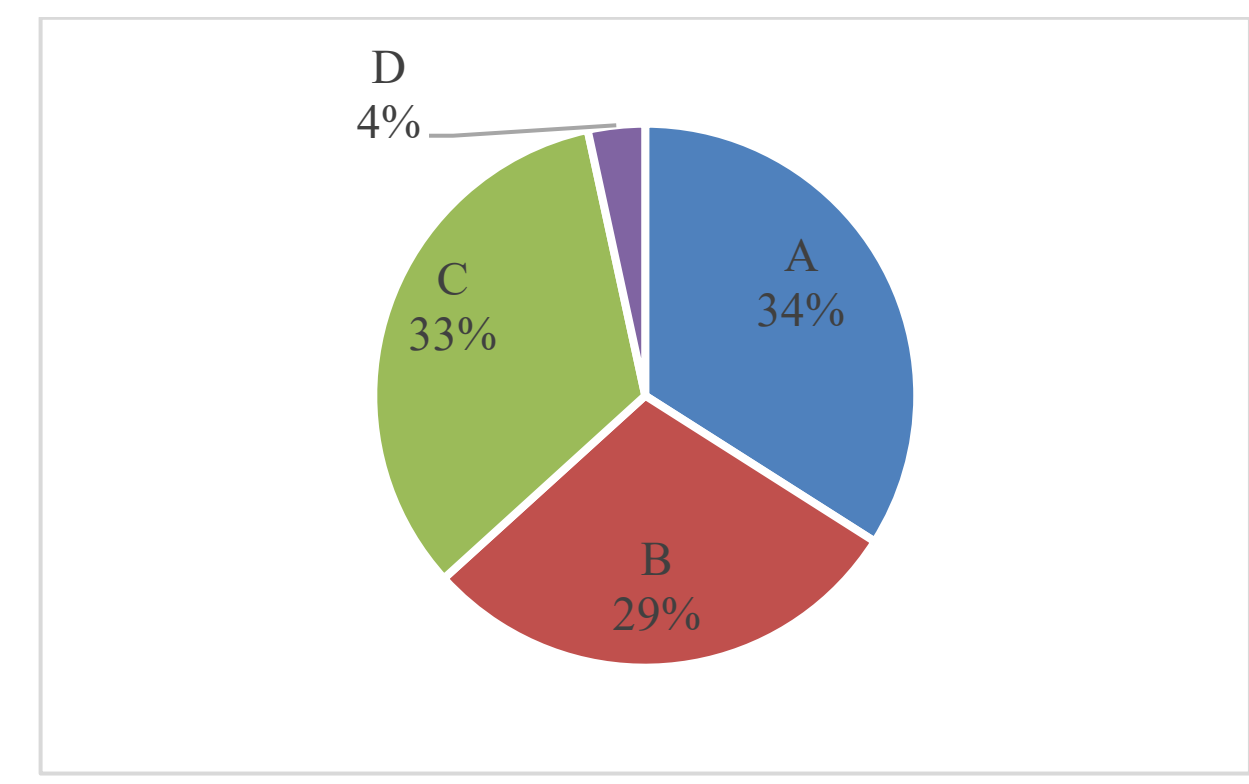

Fig. 2. Response to Q2

3. **From a technical perspective and to the best of your knowledge, what are the main drivers for applying GFM technologies in the country or markets? (Multiple choices).**
   A. There is currently no sign of any need for GFM technology in my country/market.
   B. Act as the only voltage source to form the voltage and frequency of the power network, e.g., islanded power network or microgrid.
   C. Improvement of inertia.
   D. Improvement of voltage stiffness under the weak grid
   E. Improvement of fault current level.
   F. Others (Please specify if you don't agree with all above options):

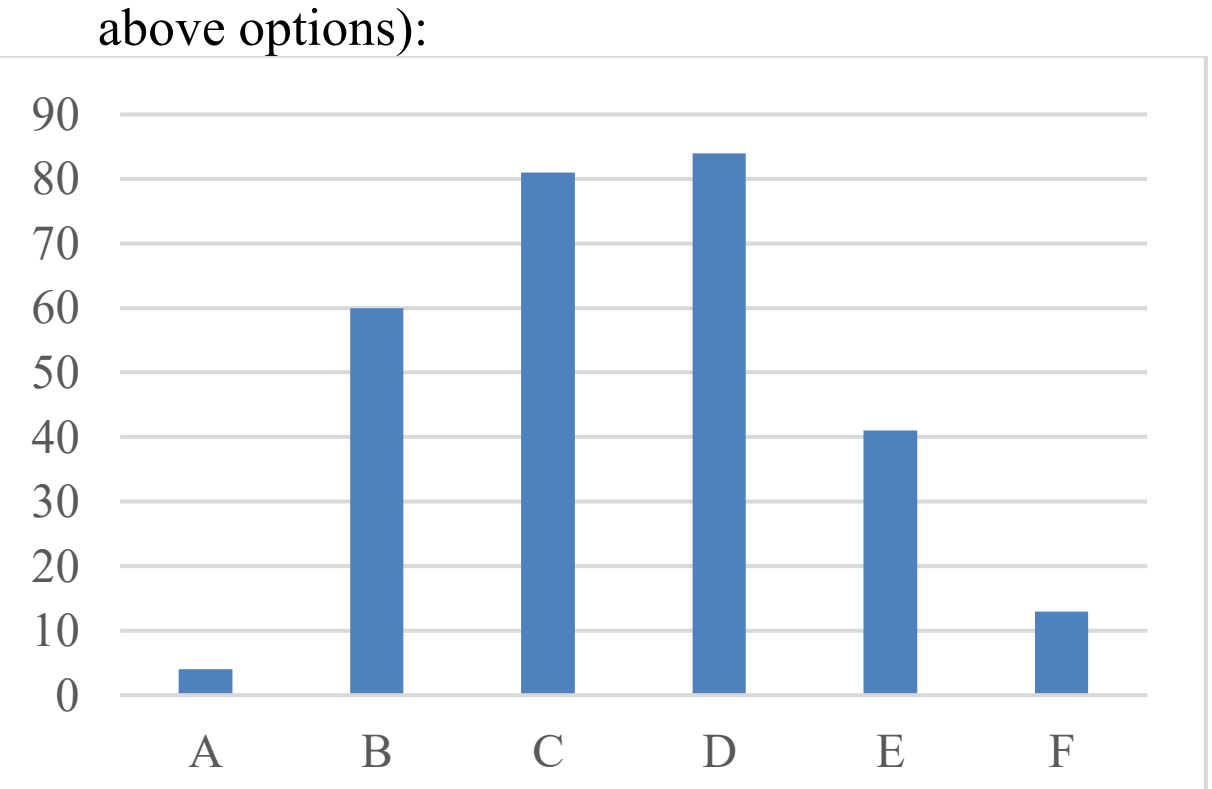

Fig. 3. Response to Q3

4. **There are two major types of GFM technologies, i.e., machine-based [synchronous condensers (SC)], and converter-based (GFM-CBR). Given the fact that SC is already very mature and can provide high fault current which CBR cannot. What is the motivation of developing/deploying solutions based on GFM-CBRs? (Multiple choices).**
   A. Currently SC fulfills all demands in our country/market, and there is no interest in developing/deploying CBR-based GFM solutions.
   B. GFM-CBR may become the preferable solution due to technical factors, such as the programmable nature of GFM-CBR controller offering greater flexibility and enabling them to provide large and adjustable inertia/damping to the system, while the inertia/damping of SC is usually fixed and low.
   C. GFM-CBR may become the preferable solution due to

operational factors, such as the maintenance of GFM-CBR is much easier due to the absence of rotational components.

D. GFM-CBR may become the preferable solution due to project development factors, such as the transportation of heavy SC machinery to remote areas with weak grids is very costly or even impossible, while transporting (modular) GFM-CBRs is much easier.

E. In the specific application that requires both longer energy buffer with energy storage system (ESS) and voltage stiffness improvement, adopting GFM-ESS is a better/cheaper solution compared with that using grid following ESS + SC

F. The choice is exclusively driven by economic factors: both solutions can be considered equivalent as long as they fulfill the requirements. The one with the lower total cost will be adopted.

G. Others (Please specify if you don't agree with all above options):

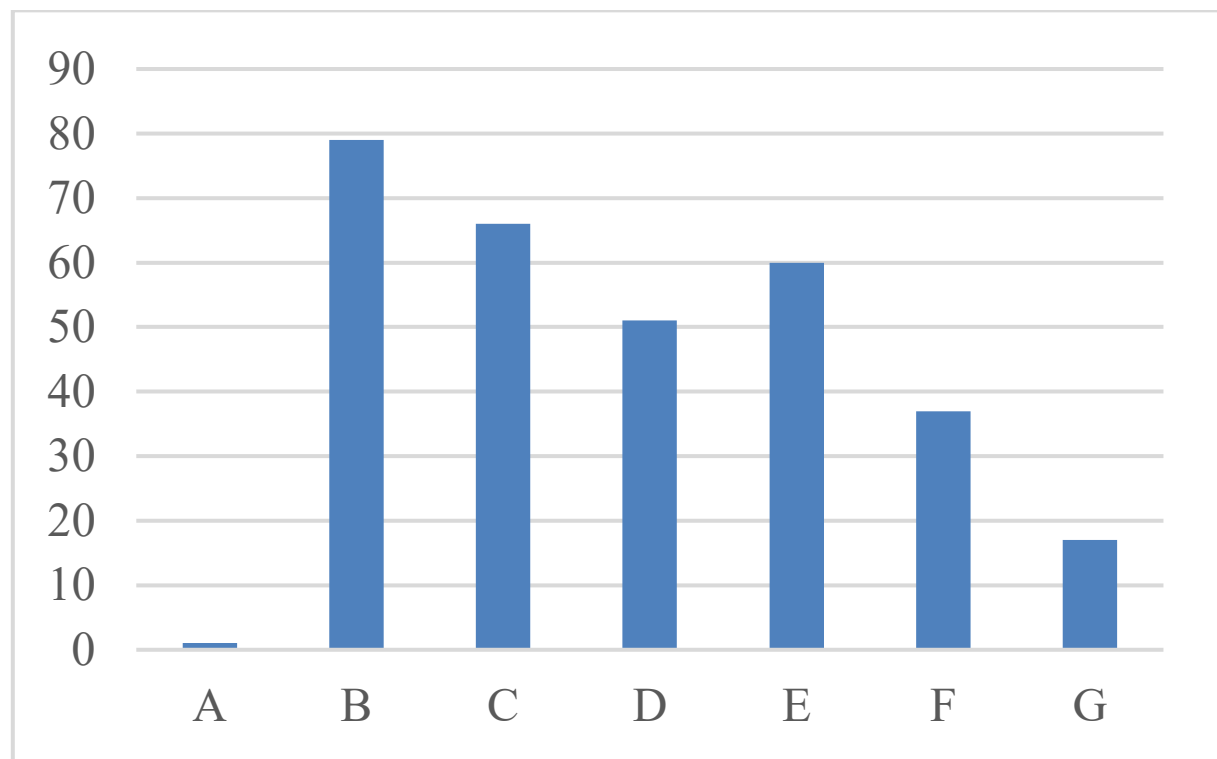

Fig. 4. Response to Q4

TABLE III
RELEVANT COMMENTS TO Q4

| No | Comments |
|---|---|
| 1 | By combining the advantage of both technologies, where GFM can provide black start and some reactive power for damping of overvoltage after the fault clearance and the SynCon provide high fault level contribution and MVArs. |
| 2 | Note that SC cannot damp active classical active power oscillation which may be re-introduced with the application of purely SC based solutions. also an eye on the SSTI characteristics should be maintained- many SC are SC via flywheel and the torsional vulnerabilities are unclear at this time. SC may be in the presence of legacy Grid follow type 3 wind etc |
| 3 | GFM-CBR may become the preferable solution due to environmental benefits if the energy buffer is implemented by a CO2 free technology |
| 4 | Many BESSs are awaiting to connect to Australia grid anyway, so we prefer simply changeing their software to make them GFM |
| 5 | "Basic" GFM can be implemented in existing hardware of e.g. OWF, solar-PV and HVDC. This comes at a very limited extra cost |
| 6 | In certain pockets, machine-based solutions may be providing too much fault current. GFM-CBR may be preferable in that adjustable inertia and voltage control is provided while limiting fault current to 1 pu. |
| 7 | The fundamental difference between a SC and an CBR is that a SC is an ancillary device in the system while an CBR is a generating unit. The whole idea of green transition is to increase the penetration of renewable CBRs in the system instead of the number of ancillary devices like SC, simply because it is renewable CBRs that produce the real power that are consumed by customers. Therefore, it is a more social economic beneficial solution if the CBRs themselves can maintain the system stability without introducing extra ancillary devices into the system, which is extra cost for the whole society. GFM control provides such a possibility that by having GFM CBRs in the system, we can potentially avoid adding SCs into the system, which is social economic optimal. However, the penetration of renewable CBR in the system cannot be increased by purely applying GFL-CBRs without using ancillary devices like SC to maintain system stability |
| 8 | Also apply to RES and HVDC, GFM control in an already planned facility is always cheaper than installing dedicated assets |

**5. There are many types of GFM-CBR technologies, e.g., GFM-ESS, GFM-HVDC/FACTS, GFM-Wind/PV. What are the unique benefits of GFM-ESS? (Multiple choices).**

A. There is no clear benefit of GFM-ESS compared with other GFM-CBR technologies.

B. GFM-ESSs provide an inherent energy buffer that can offer substantial transient active power support and sustain it over longer timescales, typically through high inertia.

C. GFM-ESSs, with their inherent energy buffer, can also participate in more energy-intensive services such as fast frequency response (FFR), frequency containment reserves (FCR) and eventually black start, offering both technical and economic advantages.

D. GFM-ESSs can control the power in both directions (absorbing/supplying), giving them a wider range of support.

E. Others (Please specify if you don't agree with all above options):

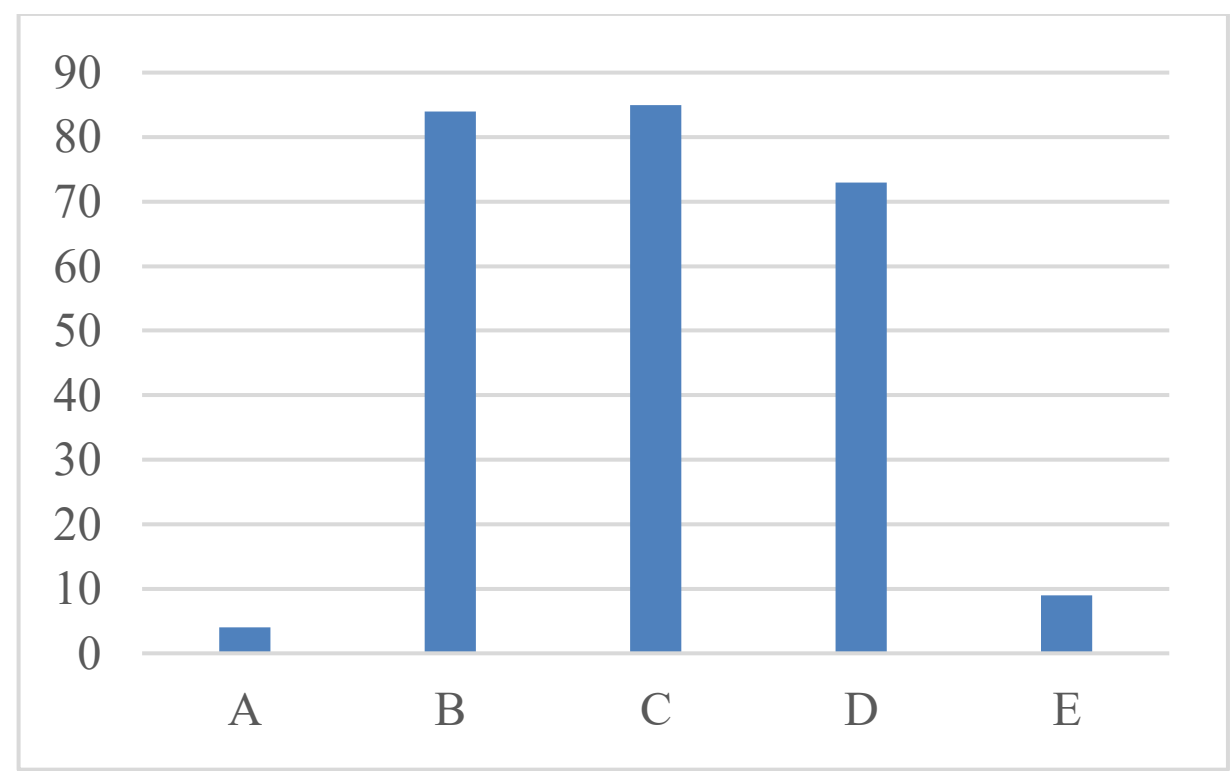

Fig. 5. Response to Q5

**6. Are there any real-world GFM-ESS projects in your country? (Multiple choices possible).**

A. Yes. The project is finalized and under operation.

B. Yes. The project is in the progress of execution.

C. Yes. The project is under the planning stage (decided

or in tender phase).

D. The project is equipped with GFM functions, but is not enabled yet.

E. There are relevant discussions, but no decisions have been made.

F. No.

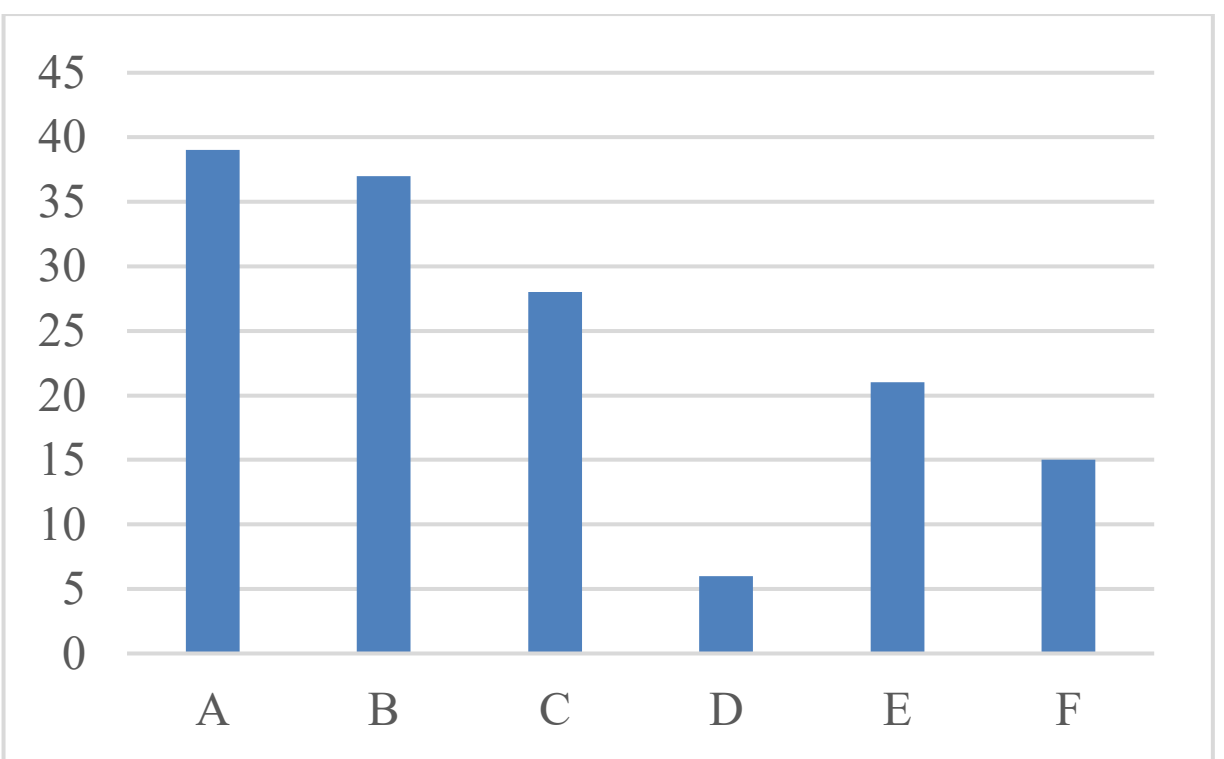

Fig. 6. Response to Q6

**7. Are there any grid-following (GFL)-ESS projects in your country that will be modified to GFM-ESS? (Multiple choices possible).**

A. Yes. The project is finalized and under operation.

B. Yes. The project is in the progress of execution.

C. Yes. The project is under the planning stage (decided or in tender phase).

D. There are relevant discussions, but no decisions have been made.

E. No.

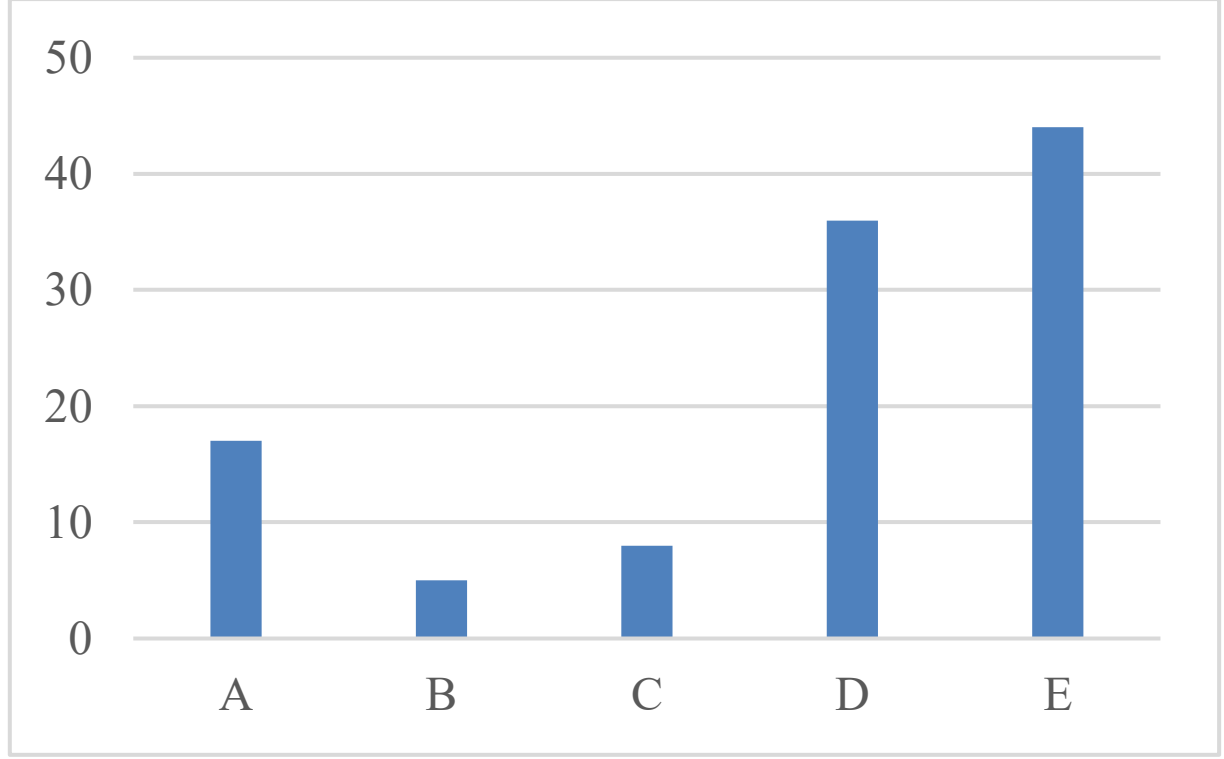

Fig. 7. Response to Q7

**8. For Transmission System Operators (TSOs), please select 3 most challenging issues for adopting GFM-ESS (3 choices).**

A. Various design and control behavior from different vendors.

B. Lack of suitable methodologies and associated tools, including simulation benchmark and models, to assess the needs and effectiveness of GFM-ESS during the planning stage (e.g., determining optimal rating and placement of GFM resources)

C. Lack of suitable methodologies and associated decision support tools to ensure the effective and timely deployment of GFM resources at the operational stage.

D. New small-signal stability risk introduced by GFM-ESS due to its interaction with GFL-CBRs, synchronous generators (SGs), etc.

E. Uncertainties about the dynamic behavior of GFM-ESS under grid faults (e.g. prioritized reactive current injection, nature fault current injection, slowdown of the power recovery after fault, etc), which brings in challenges for system-level transient stability analysis and coordination with protective relay.

F. It is not clear how GFM's performance will be compromised when limits are reached (current limit, energy limit, DC voltage limit, etc), and the impact on the system.

G. Regulatory barriers to develop/deploy GFM-ESS on the TSO side and lack of incentives on the grid user side.

H. Others (Please specify if you don't agree with all above options):

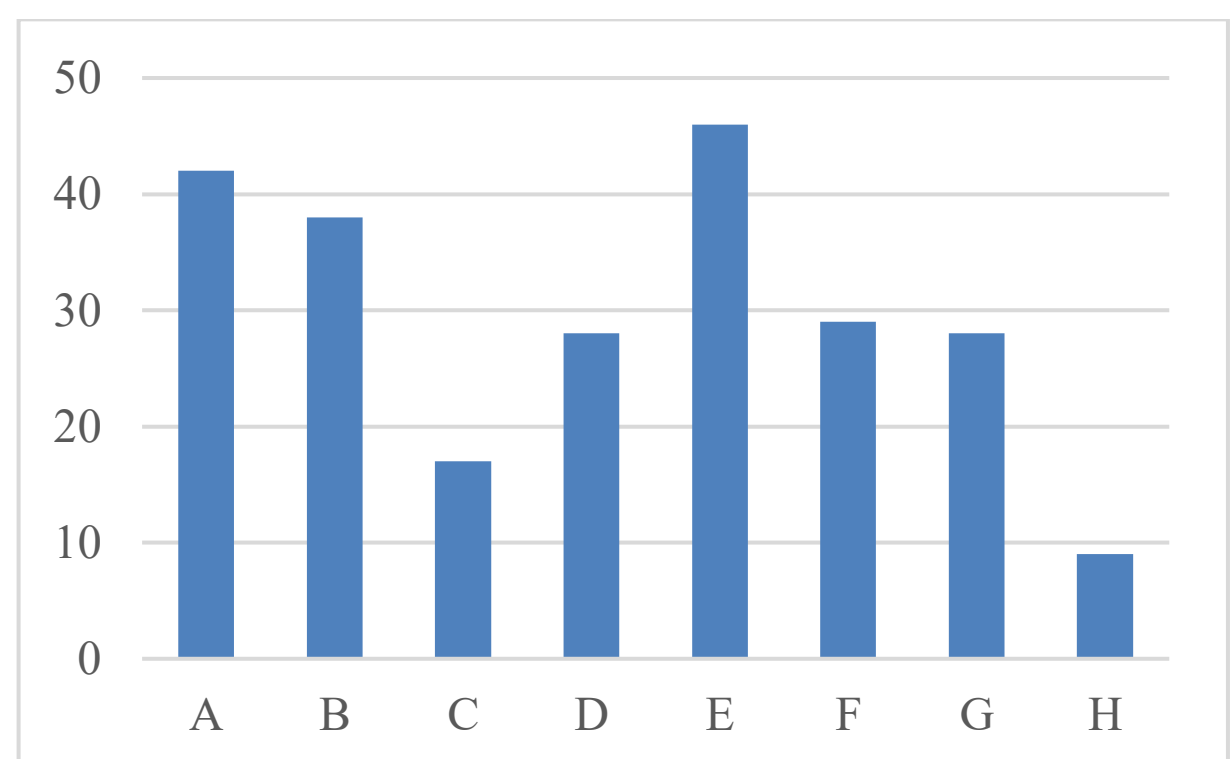

Fig. 8. Response to Q8

**9. For power-plant owners/developers, please select 3 most challenging issues for adopting GFM-ESS (3 choices).**

A. The fault ride through behavior (including transient overload capability) of GFM-ESSs from different vendors is not standardized.

B. The energy market is not updated by considering the ancillary services provided by GFM-ESS.

C. No clear incentives, compensation, or revenue stream to motivate developers or plant owners to choose GFM over GFL for services that only GFM-ESS can provide.

D. No clear connection process/grid codes for GFM-ESS projects (in terms of performance standard and tests expected to be passed) as compared to more established and clear connection process for GFL-ESS projects.

E. No clear method yet to quantify some of the services GFM provide (e.g. stiffening voltage waveform or providing inertia contribution, etc.).

F. New small-signal stability risk introduced by GFM-ESS due to its interaction with GFL-CBRs and SGs, etc.

G. Missing field experience and concerns about lifetime.

H. Vendors feel difficult to fulfill the technical requirements raised by power plant owner/developer,

e.g., the requirement of dynamic change of Inertia level (flexible GFM performance), which blocks the future stability market.

I. Others (Please specify if you don't agree with all above options):

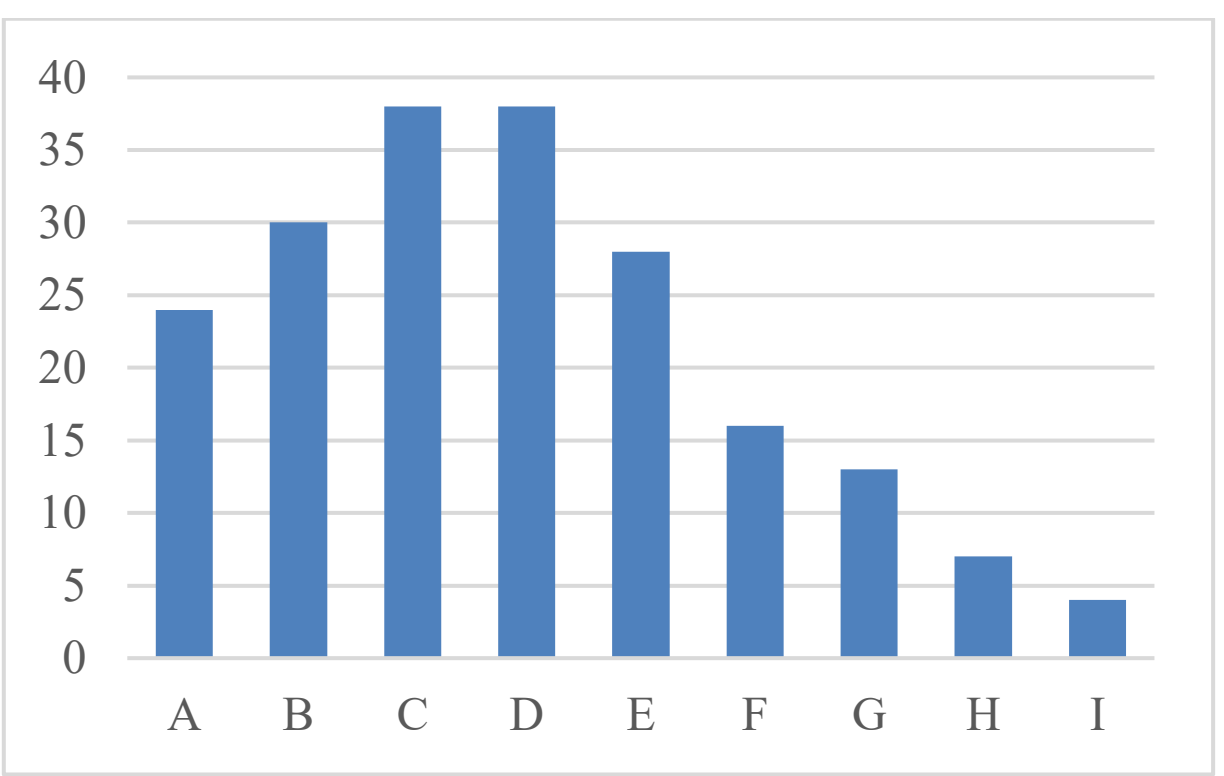

Fig. 9. Response to Q9

**10. For original equipment manufacturers (OEMs), please select 3 most challenging issues for developing GFM-ESS (3 choices).**

A. No clear connection process for GFM-ESS projects (in terms of performance standard and tests expected to be passed) as compared to more established and clear connection process for GFL-ESS projects.

B. Various design and performance requirements from different buyers.

C. Some requirements of GFM-ESS are not realistic (e.g., requiring positive damping at fundamental frequency).

D. Some requirements of GFM-ESS could significantly increase the cost (e.g., requiring higher transient overloading capability), which is not fully recognized by some system owners.

E. Long lasting certification process.

F. Uncertainties of the market evolution for GFM solutions, which creates the lack of incentives for product development.

G. Operating GFM-ESS requires high power ramp-rates, which could lead to an increased aging of the battery storage system.

H. Others (Please specify if you don't agree with all above options):

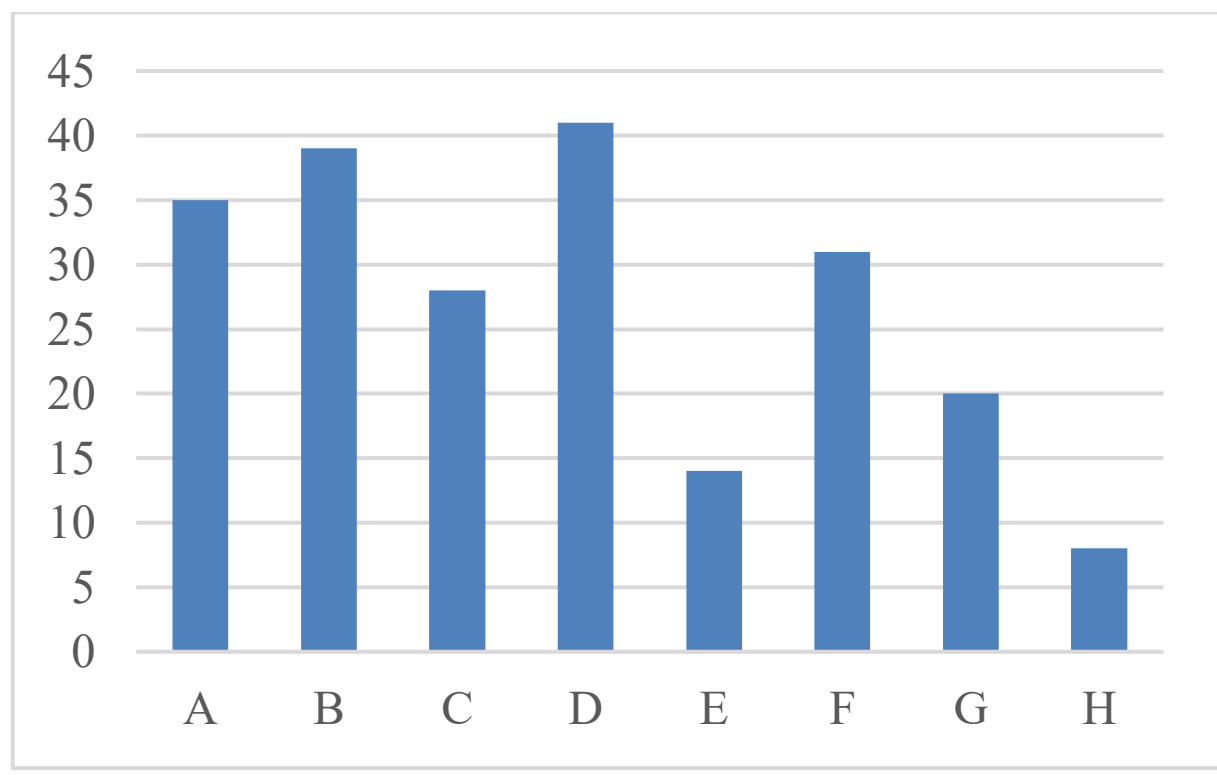

Fig. 10. Response to Q10

**11. What need to be further developed/investigated for accelerating the application of GFM-ESS? (Multiple choices)**

A. Benchmark RMS/EMT model of GFM converters.

B. Benchmark grid-model including active loads for performance verification of GFM-converters.

C. Grid code/technical specifications, as well as compliance testing guidelines.

D. New energy market with the consideration of ancillary services provided by GFM-ESS.

E. Interaction analysis between GFM-, GFL-ESSs, SGs and other equipment, multiple GFM-ESS plants.

F. Coordination between the fault ride through of GFM converters and power system protection.

G. Methodologies for optimal sizing, as well as identifying the optimal locations of GFM-ESS.

H. Considering changing the firmware of some existing GFL ESSs to GFM with minimal cost (just to provide some core GFM services, without expecting much overload capability).

I. Others (Please specify if you don't agree with all above options):

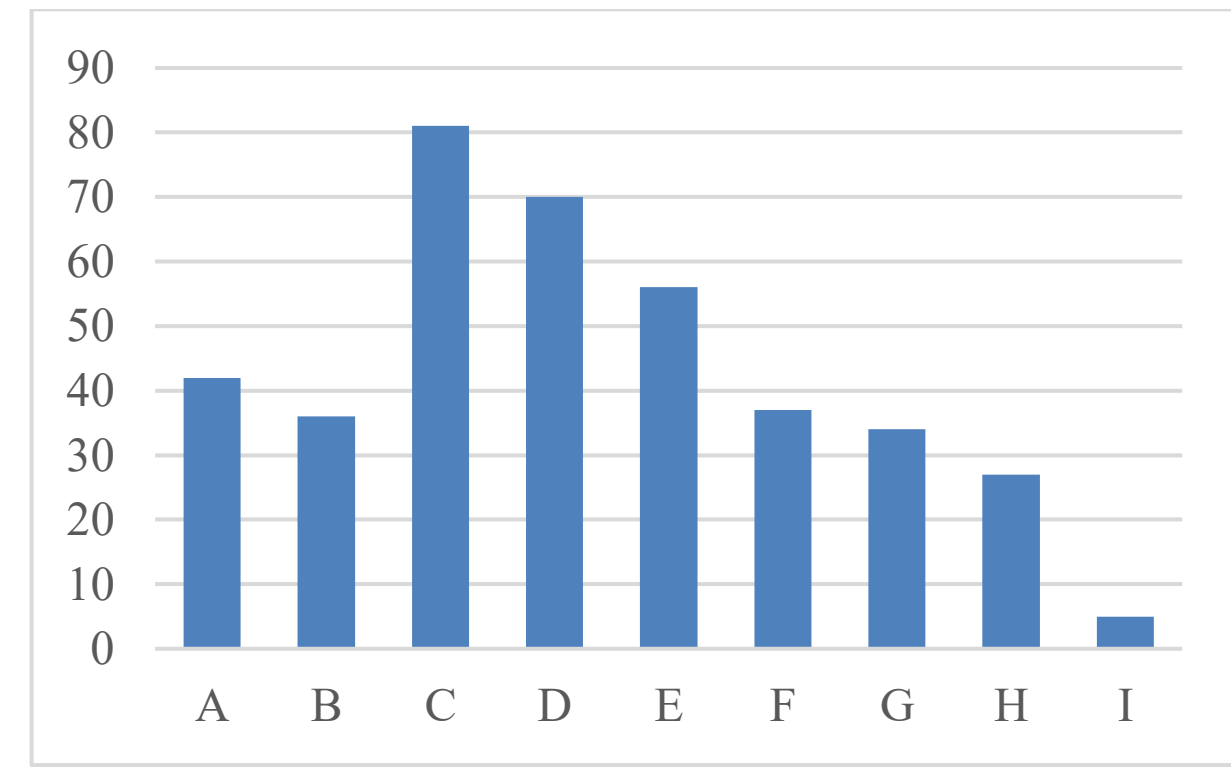

Fig. 11. Response to Q11

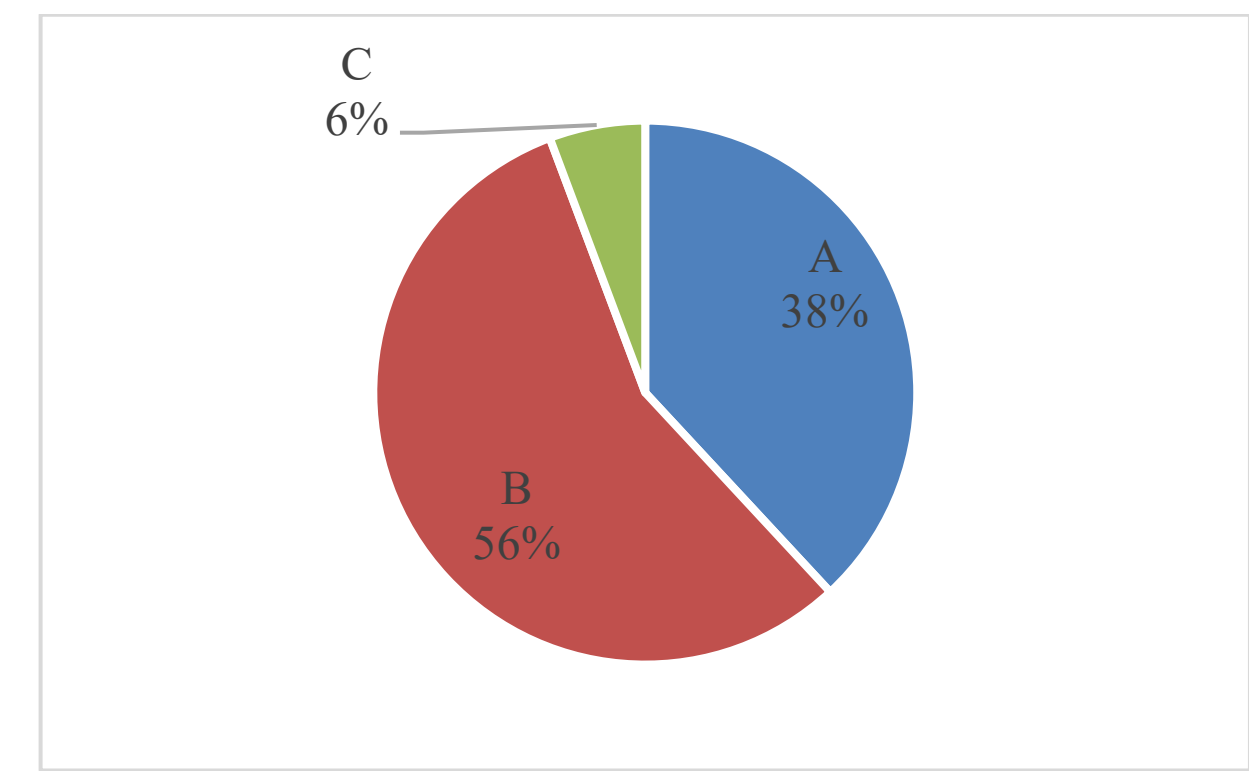

Fig. 12. Response to Q12

**12. Do you see the necessity of standardizing the transient overloading capability of GFM-ESS ? (Single choice)**

A. No, since the optimal transient overloading capability is project-specific.

B. Yes, while the optimal transient overloading capability might be project-specific, make it standardized can simplify the design and analysis of GFM-ESS from both OEMs and TSOs perspective.

C. Others (Please specify if you don't agree with all above options):

**13. Do you think the high transient overloading capability of GFM-ESS is essential to the reliable operation of the protective relay? (Single choice)**

A. Yes, the proper operation of overcurrent relay relies on the high fault current magnitude.
B. No, as the distance and differential protection are more widely used in transmission system compared with overcurrent relay, whose operation do not rely on the fault current magnitude.
C. Others (Please specify if you don't agree with all above options):

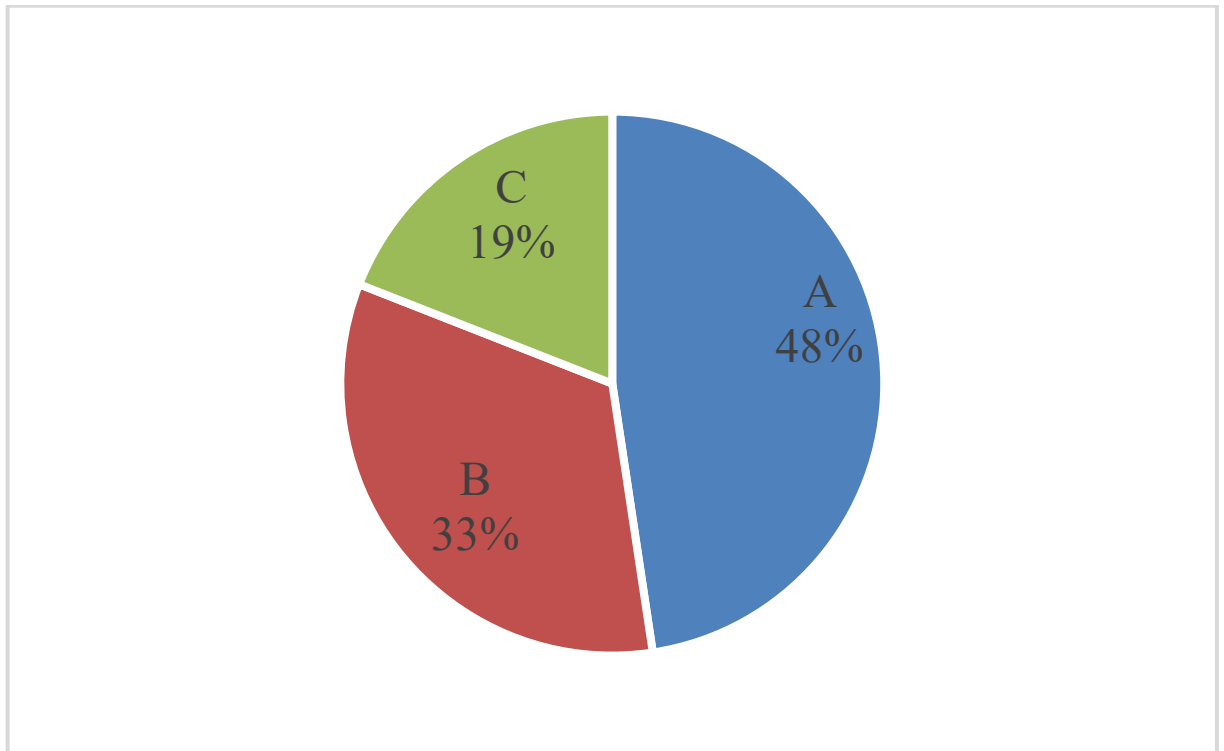

Fig. 13. Response to Q13

TABLE IV
RELEVANT COMMENTS TO Q13

| No | Comments |
|---|---|
| 1 | No. as its a cumulative effect of how many GFM you have and what their pre-fault operating point is as to how much useable overload current exists for a given condition. Its entirely possible during high concentration of CBR with low MW yield you end up with at least as much as before; and potentially too much in some cases. the major benefit to distance relays is the clear polarity and clear basis to introduce calculations of settings. |
| 2 | There is a gap in the protective relay technology develop to adapt to CBR integration. Currently, CBR need to increase overload to adapt existing overcurrent relays. . |
| 3 | As we have previously seen with CBRs, other aspects are very important, such as the negative sequence behavior. |
| 4 | I tend to choose point 1 because even distance protection requires higher current to enter the "reach" zone |
| 5 | No, a high transient overloading capability of all GFM-ESS is not essential to the reliable operation of the protective relays. Ensuring an adequate fault current level for protection depends on the protection strategy of the grid operator. In case a minimum fault current level is necessary, it needs to be maintained. There are multiple options how the grid operators can maintain a minimum firm fault current (i.e. by own grid assets or market-based). |

**14. What is your opinion on improving the transient overloading capability of GFM-ESS to enhance its support to the power network during the fault and after the fault recovery? (Single choice)**

A. This is one of the essential benefits of the GFM-ESS and is critical to the power system. Therefore, high transient overloading capability of GFM-BESS is demanded even at the expense of high cost.
B. It would be beneficial but is not the essential demand of the power system. Therefore, high transient overloading capability of GFM-BESS is not attractive considering it requires higher cost.
C. Others (Please specify if you don't agree with all above options):

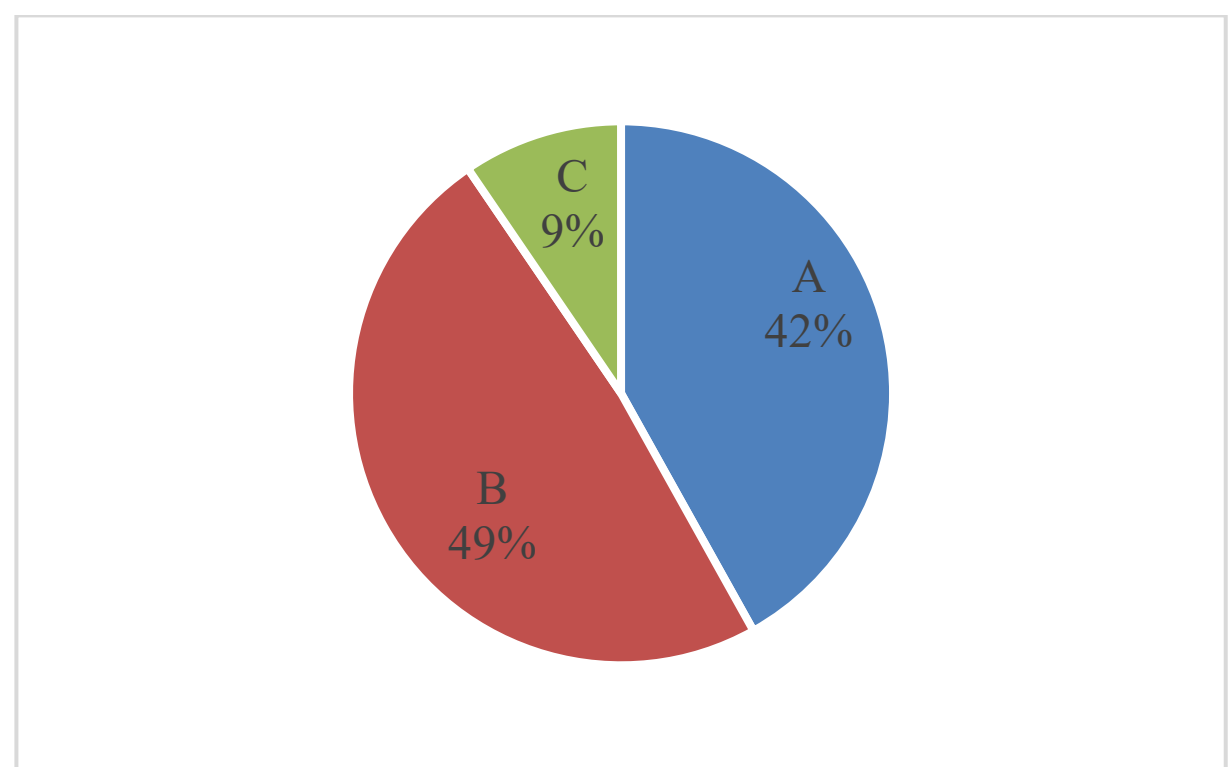

Fig. 14. Response to Q14

TABLE V
RELEVANT COMMENTS TO Q14

| No | Comments |
|---|---|
| 1 | It is an economic decision relative to the market incentive for providing it. |
| 2 | Might be essential if a performance similar to synchronous machine is expected |
| 3 | Compared to conventional SC technology, where overload is a byproduct of power conversion technology, the realization of transient overload capability with converter-based technology is not for free. Converters must be designed for it. Such a design typically has an impact on costs but it can also provide value. For maintaining minimum levels of grid supporting power not necessarily all GFM-ESS need to have transient overload capabilities. Transient overload capability for GFM-ESS and other CBRs should be market-driven and voluntary, but not mandatory. |

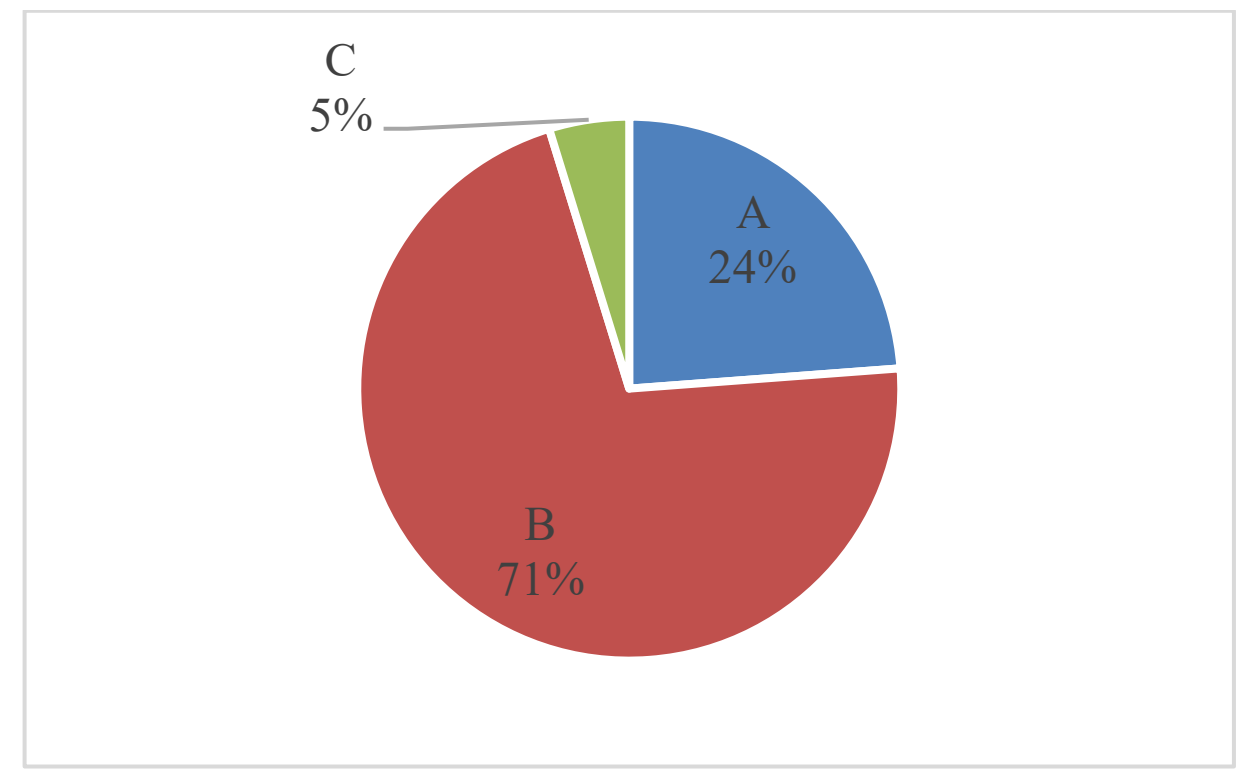

Fig. 15. Response to Q15

**15. What is your opinion on the adequate requirement of reactive current profile of GFM-ESS during grid faults? (Single choice)**

A. The GFM-ESS should prioritize reactive current injection during grid faults, which is similar to that of GFL-ESS.

B. The GFM-ESS should provide a nature response of voltage source during grid faults, except the current magnitude is limited. Therefore, the injected reactive current will be dependent on fault conditions and will not be prioritized by a control action.

C. Others (Please specify if you don't agree with all above options):

**16. The battery-based GFM-ESS might introduce a more frequent charging and discharging of the battery due to the provision of ancillary service like inertia, which might accelerate the ageing of the battery. What is your opinion on this issue? (Single choice)**

A. The inertia service only costs small cycles of batteries, thus its impact on degradation is negligible. The major impact is from other ancillary services, like FRR, FCR, etc. But these ancillary services can also be provided by GFL-ESS, and GFM does not make a difference on these ancillary services except the inertia service.

B. The inertia service although leads to small cycles time (a few seconds) and energy wise, but if transient over-loading is considered, it can considerably accelerate the ageing of the battery.

C. The number of events (large and small signal) that the GFM-ESS is exposed to, shall be specified to enable the assessment of the lifetime.

D. Others (Please specify if you don't agree with all above options):

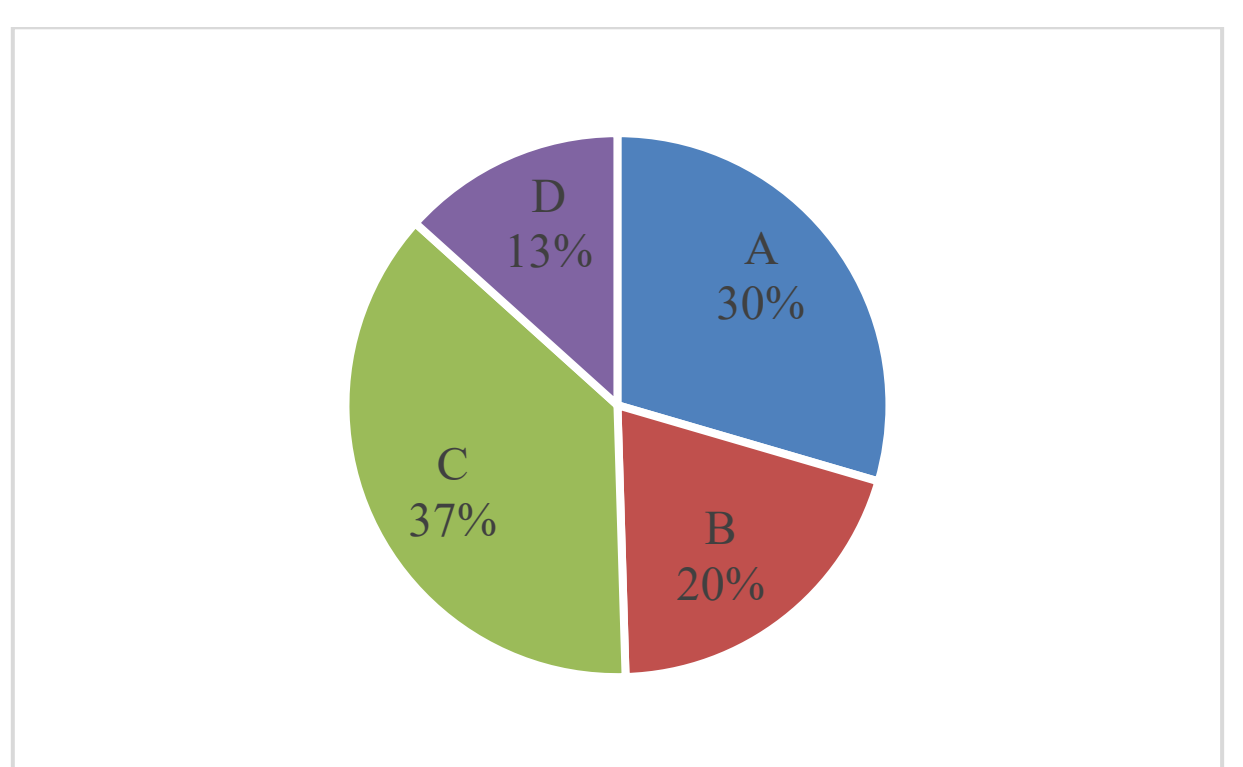

Fig. 16. Response to Q16

TABLE VI
RELEVANT COMMENTS TO Q16

| No | Comments |
|---|---|
| 1 | OEMs should develop degradation models and cooperate with TSOs to analyze the expected inertial response, and devise appropriate market services that remunerate owners taking into consideration the marginal cost of the response |
| 2 | The aging of energy or battery storage depends on multiple factors like the storage technology and also on the load profile, that depends on one side on the local disturbance conditions and on the other side on the stability service extent and compensation objectives of a GFM ESS (i.e. how much inertia power or damping power and other services it should provide, which is determined by its parametrization). |
| 3 | In our experience the small depth of the cycles cause by inertia, and small number of cycles caused by FRR, FCR, etc. causes negligible degradation of the battery. |

## III. STATE-OF-THE-ART AND CONSENSUS

This section summarizes the state-of-the-art and consensus emerging from the survey responses. By analyzing the collective input from academia, TSOs, OEMs, developers, and consultants across the globe, several important trends and agreements become evident, which are summarized as follows:

### *A) Definition of GFM Behavior*

Historically, GFL and GFM-CBRs have been differentiated based on their dedicated control schemes: GFL-CBRs rely on a phase-locked loop (PLL) to synchronize with the power grid, whereas GFM-CBRs are characterized by the removal of the PLL, and adopting power-based synchronization instead. This control-loop-based distinction has been accepted for quite some time in the past. However, in recent years, there is a growing consensus that GFM-CBR should be defined by its dynamic behavior, i.e., a voltage source behind an impedance, without specifying how specific control loops are implemented internally. As pointed out by recent literatures, the presence or absence of a PLL is no longer a definitive discriminator between GFL- and GFM-CBRs. For instance, the hybrid synchronization control (HSC) of GFM-CBR is proposed in [14]-[15] by integrating PLL with the power-based synchronization control, which improves the stability robustness of GFM-CBRs. Moreover, ABB has demonstrated that GFM behavior can also be realized using PLL only with the vector current control (i.e., without using the power-based synchronization), provided that the PLL and alternative voltage control (AVC) are properly designed to guarantee the dynamic features of a voltage source [16].

This growing consensus for the feature-based definition of GFM-CBRs is also reflected in responses to Q1. As shown in Fig. 1, most respondents believe that GFM functionality should be defined based on the set of dynamic behavior it provides, rather than a specific control structure. Specifically, 68% of respondents selected option A: "*GFM refers to a specific set of features that characterize the transient response of a grid-connected device…*", whereas only 5% selected option C: "*GFM refers to the removal of the classical PLL synchronization in CBR control…*"

This evolution from a control-based definition to a feature-based one has significant implications on grid codes, compliance testing, and future development of GFM technologies. It suggests that the industry is moving towards a more performance-oriented framework, where the focus is on what a device does for the system, rather than how its control loops are implemented internally. This aligns with the practical needs of TSOs, who require predictable and beneficial grid responses, and OEMs, who need the flexibility to innovate in their control designs to meet these performance targets.

*B). Motivation for Deploying GFM-CBRs Over Synchronous Condensers*

Synchronous condensers (SCs), which also behave as the voltage source behind the impedance, is the mature solution for improving system strength. Moreover, SCs can deliver much higher short-circuit currents than CBRs. Hence, one might question the motivation for developing and deploying GFM-CBRs over SCs. This concern is raised via Q4 of the survey. From the response, it can be found out that option B, C, D, E are most selected. Together with the relevant comments given in Table III, it indicates that there are multiple factors—technical, economic, and social—that drive the adoption of GFM-CBRs over SCs, which are detailed as follows:

*1) Technical benefits*:

*Better damping*: SCs feature inherently low damping. As pointed out in comment 2 of Table III, SCs can hardly damp power oscillations and may even introduce sub-synchronous torsional interaction (SSTI) vulnerabilities, especially in flywheel-based designs. In contrast, the damping of GFM-CBRs is programmable and can be tuned to be much larger than that of SCs, as indicated in option B. This advantage has been verified by Energinet (the Danish TSO) through simulation tests, in which the dynamic responses of GFM-CBRs and SCs during grid faults are compared by using vendor-specific models, as shown in Fig. 17 [17]. It can be seen that although SC provide better voltage support during the fault due to their high fault current contribution, their post-fault dynamics are highly oscillatory due to the low damping. In contrast, GFM-CBRs yield superior post-fault voltage control and damping capability, which can damp the power oscillation and restore system voltage in a much faster manner after grid faults.

*Controlled fault current*: While higher fault current is generally preferred, there are scenarios where the network already has excessive fault current but still requires GFM-type voltage support. As noted in comment 6 of Table III, such situations occur in some regions in southeast part of China where the grid is tightly interconnected with many synchronous generators (SGs), but the number of online SGs could vary significantly depending on the power generation of renewable energies. This creates a unique phenomenon: the system has sufficient (or even too much) fault current in general, yet still needs additional voltage support when few SGs remain online. In this scenario, GFM-CBRs offer an ideal solution over SCs, as they can provide voltage support while limiting fault current to, for example, 1 per unit.

*2) Benefits in project deployment and maintenance*

GFM-CBRs also offer distinct advantages during project deployment and maintenance phases, as reflected in strong support for options C and D. For instance, the modular design of GFM-CBRs considerably eases their transportation to remote areas with weak grid interconnections compared to bulky SCs. Likewise, the absence of rotational components in GFM-CBRs simplifies maintenance requirements and reduces associated operational costs. As pointed out in [18], the maintenance cost of SCs is around 2.5 times of GFM-CBR with similar power ratings.

*3) Socio-economic benefits*

The fundamental objective of the green transition is to increase the penetration of renewable energy sources. Consequently, a large number of renewable generation units, along with battery storage systems, are planned to connect to the power grid via CBRs. In such a case, it is a more socio-economically beneficial solution if CBRs themselves can maintain system stability without introducing extra ancillary stabilizing devices such as SCs. As highlighted in option E and comments 3–5, 7–8 of Table III, implementing GFM control within existing CBRs, as well as CBRs that will connect to the power network in the future, is far more cost-effective than adding dedicated stabilizing devices like SCs.

Finally, as pointed out in comment 1 of Table III, GFM-CBRs can also be combined with SCs to form a hybrid solution that harnesses the advantages of both technologies, thereby providing damping, inertia, as well as high fault currents to stabilize the system. A more detailed elaboration of this hybrid solution can be found in [19].

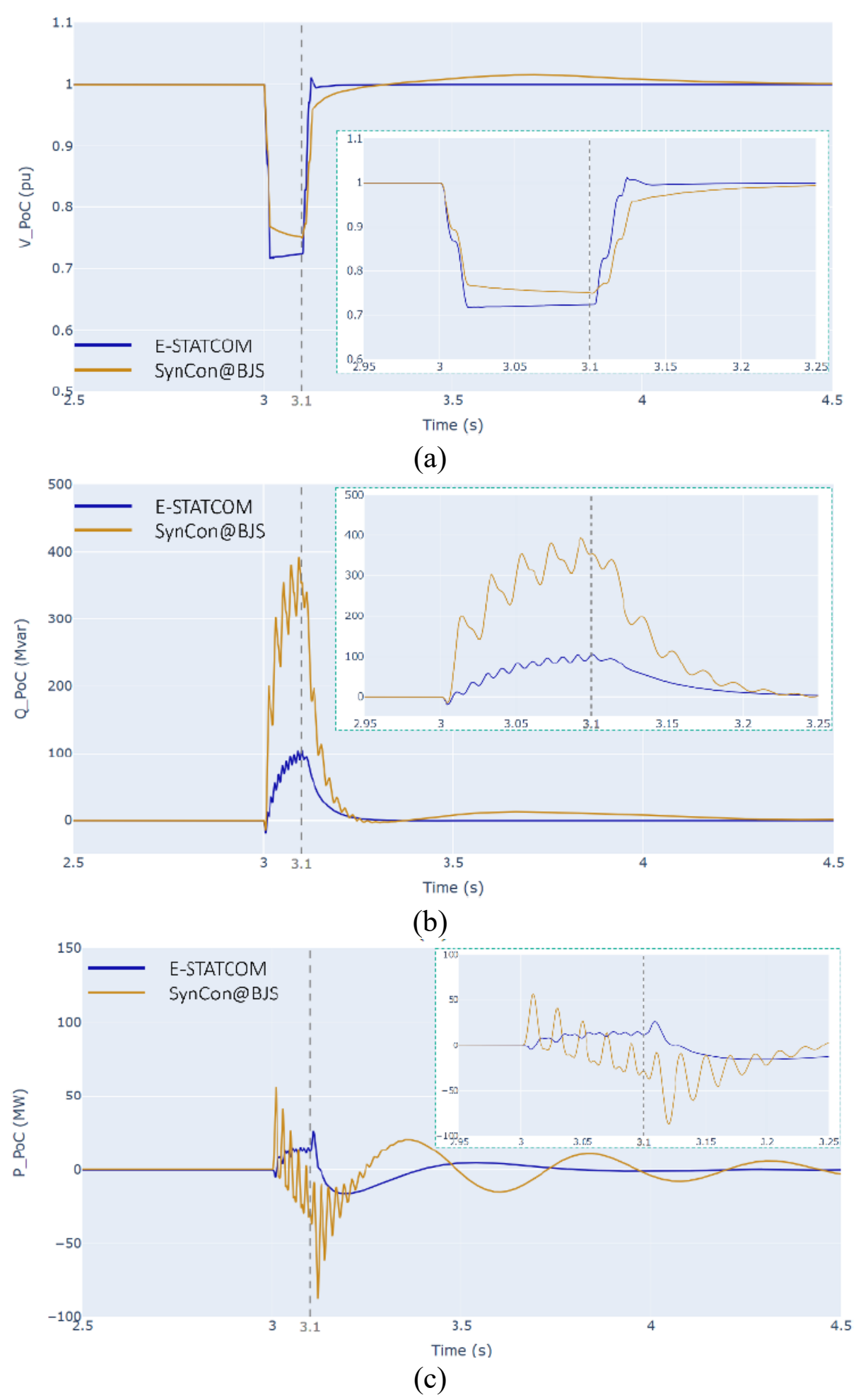

Fig. 17. Comparison of dynamic responses between synchronous condensers and GFM-STATCOM (E-STATCOM), Source: [17]. (a) PoC voltage. (b) Reactive Power. (c) Active Power.

*C) Current Injection Profile under Grid Faults*

Existing grid codes have well-established requirements for fault ride through (FRT) behavior of GFL-CBRs: they must prioritize reactive current injection during grid faults to support grid voltage [20]. This approach relies on fault detection mechanisms within the control system, followed by the active modification of the current reference. However, this control-induced current injection inevitably introduces a delayed response, resulting in delayed voltage support during grid faults and transient overvoltage issues upon fault recovery, as illustrated in Fig. 18(a) [21]. In the early stages of developing GFM control, it was common practice to switch GFM to GFL control during grid faults and prioritize the reactive current injection, and thus, the aforementioned FRT problems of GFL-CBRs were also inherited.

To address this challenge, the industry is moving toward a consensus that GFM-CBRs should behave as a voltage source behind an impedance and provide a natural response during grid faults, with the sole intervention being the current-vector magnitude limitation. This requirement has been clearly articulated in recent technical documents from ENTSO-E and Energinet: "*…GFM shall limit the output current only by reducing the current phasor magnitude while maintaining the current phasor angle of the unlimited phasor constant. It should be noted that active power priority or reactive power priority is not accepted….*" [22]-[23], as shown in Fig. 19. Under this philosophy, the injected active and reactive fault currents are not prioritized by a dedicated control action but are inherently determined by the fault conditions and the impedance between the source and the fault point. More specifically, if a grid fault manifests as a pure voltage sag, the natural response of a voltage source behind a highly inductive impedance is to inject predominantly reactive fault current. However, this reactive current injection is the natural response of the voltage source, requiring no fault detection or control intervention, and thus does not involve any time delays as that of GFL-CBRs, as shown in Fig. 18(b). Moreover, for grid faults that include both voltage sags and phase-angle jumps—which are more common in practice—the natural response of a GFM-CBR will include both active and reactive fault current components. This behavior is fundamentally different from a GFL-controlled fault response, which typically attempts to suppress the active current injection regardless of the phase-angle condition.

This evolving understanding is also clearly reflected in the responses to Q15, as shown in Fig. 15. Overall, 71% of respondents selected option B: "*The GFM-ESS should provide a nature response of voltage source during grid faults, except the current magnitude is limited. Therefore, the injected reactive current will be dependent on fault conditions and will not be prioritized by a control action.*", whereas only 24% selected option A: "*The GFM-ESS should prioritize reactive current injection during grid faults, which is similar to that of GFL-ESS.*"

This evolution from a prioritized, detection-based fault response to a natural, impedance-based response has profound implications. For TSOs, it promises a more predictable and inherently stable fault behavior that aligns with the physical characteristics of synchronous machines. For OEMs, it simplifies control design by removing the need for complex mode-switching logic. Nevertheless, careful design of the current-limiting control remains essential to ensure that the natural response stays within the physical limits of devices.

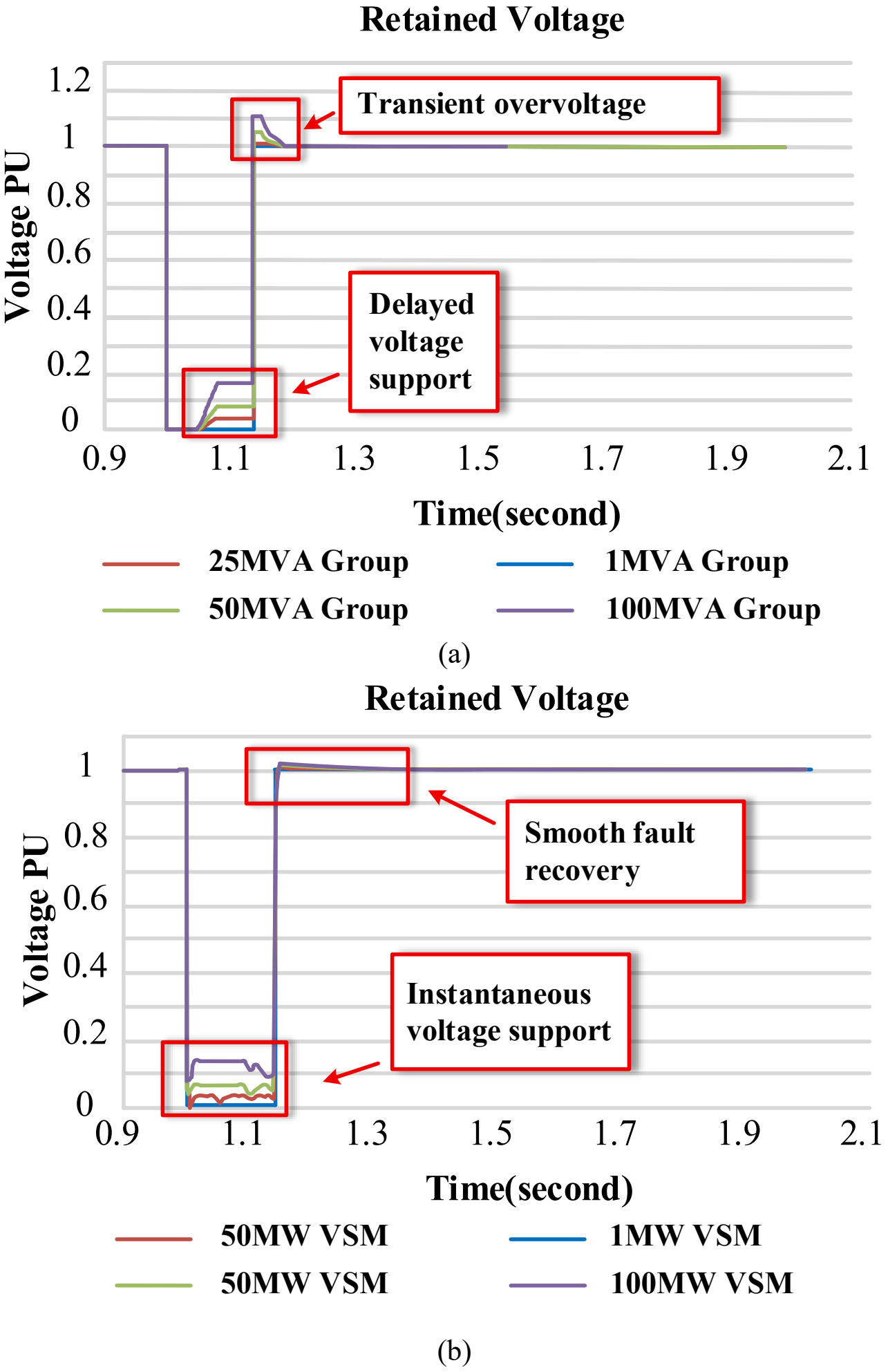

Fig. 18. Simulation results of retained voltage during FRT, source: [21]. (a) With GFL-CBRs. (b) With GFM-CBRs.

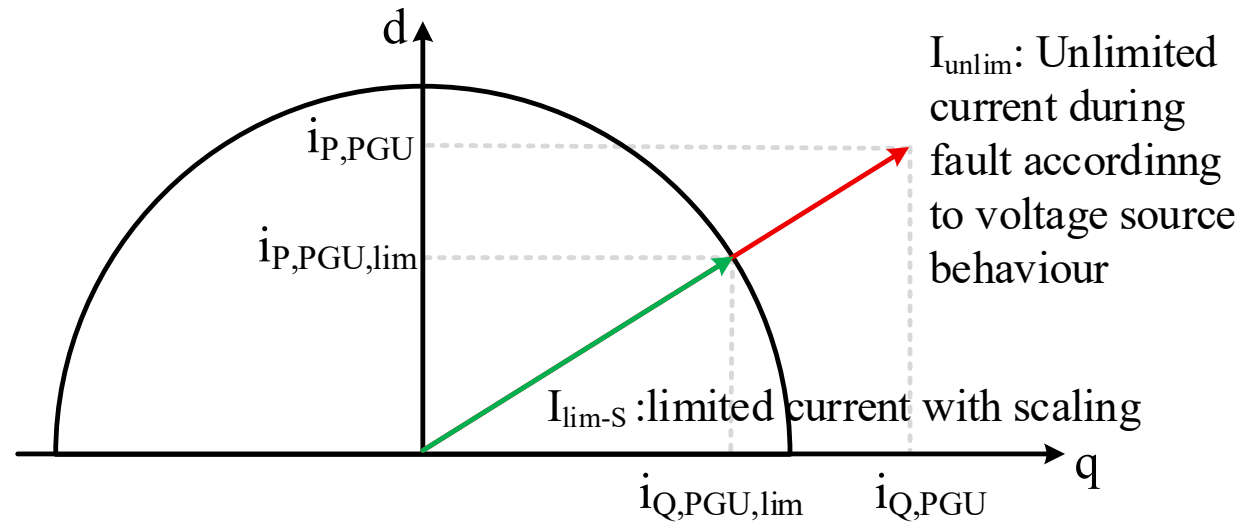

Fig. 19. Requirement of current-limitation of GFM-CBR, source: [23].

## IV. Remaining Gaps and Challenges

While previous sections have highlighted the growing consensus on various aspects for GFM technologies, the survey responses also reveal several persistent gaps and challenges that must be addressed to facilitate the widescale deployment of GFM-ESS. These challenges span regulatory frameworks, market design, technical specifications, and equipment lifetime considerations, which are detailed as follows:

### *A) Grid Code and Technical Specification*

It can be seen from the response of Q11 that option C "*Grid code/technical specifications, as well as compliance testing guidelines.*" is identified as the most important task for accelerating the deployment of GFM-ESS. While there is a growing consensus that the general requirement for GFM-ESS is to behave as a voltage source behind an impedance, providing inertia and damping to the system, the grid-code compliance testing guidelines have not been unified, which, thus, imposes significant challenges for stakeholders during the testing and certification process.

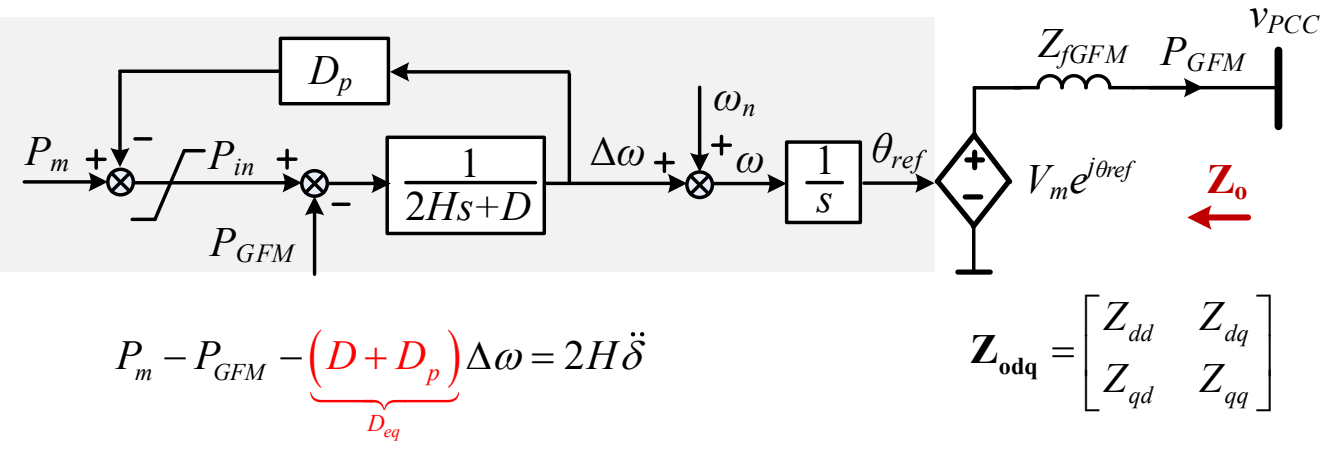

$P_m - P_{GFM} - \underbrace{(D + D_p)}_{D_{eq}}\Delta\omega = 2H\ddot{\delta}$  $\mathbf{Z_{odq}} = \begin{bmatrix} Z_{dd} & Z_{dq} \\ Z_{qd} & Z_{qq} \end{bmatrix}$

① Damping evaluation based on passivity of the output impedance matrix

$\mathbf{Z_{odq}}(j\omega) + \mathbf{Z^H_{odq}}(j\omega) > 0$

② Damping evaluation based on damping torque $D_{eq} > 0$

Fig. 20. Damping assessment based on output impedance and/or damping torque.

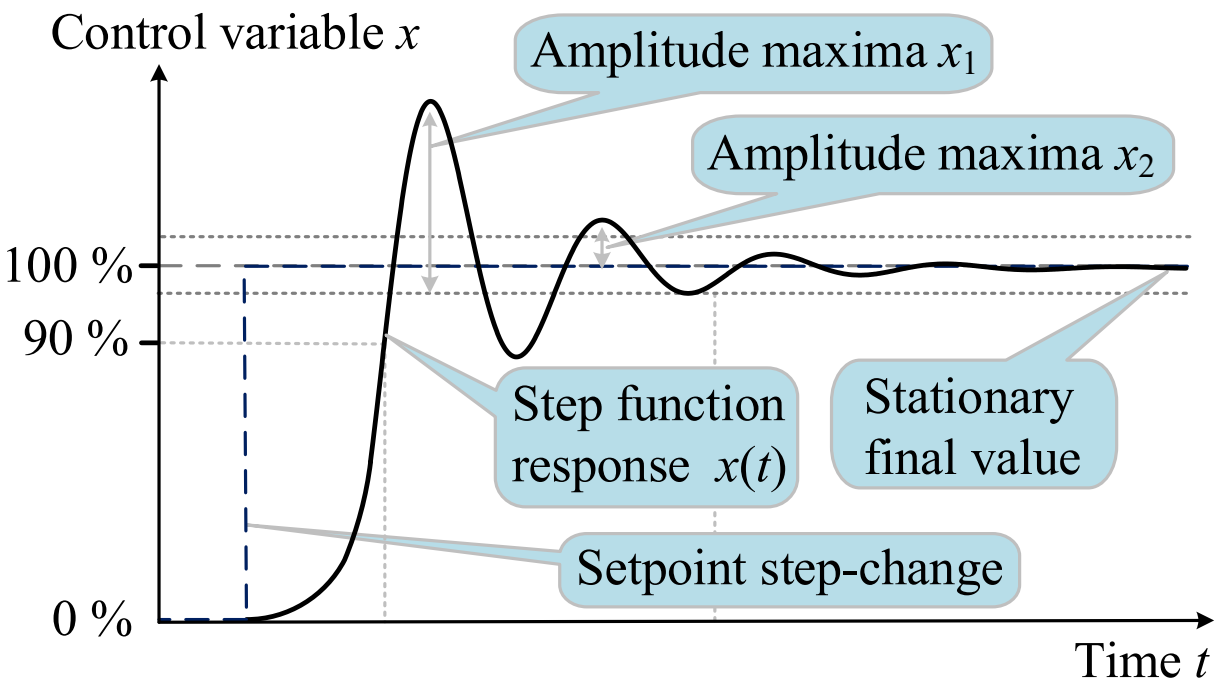

Fig. 21. Time-domain response of GFM-ESS after a step change. Source:[27]

*1) Damping*: It is widely acknowledged that GFM-ESS needs to provide positive damping to the system. However, the quantification of the damping provision varies across different grid codes and standard specifications. Some standards define damping based on the dissipative of converter output impedance [24], while others employ the damping torque analysis [25]-[26], as illustrated in Fig. 20. Recently, VDE FNN and ENTSO-E suggest calculating the damping ratio directly from time-domain response[23],[27], as shown in Fig. 21, based on which, the damping ratio $D$ can be derived as

$$D = \frac{\Lambda}{\sqrt{(2\pi)^2 + \Lambda^2}}. \tag{1}$$

where

$$\Lambda = \ln\left(\frac{x_n}{x_{n+1}}\right). \tag{2}$$

where $x_n$ and $x_{n+1}$ are defined as the differences between two consecutive peak values and the steady-state values, as shown in Fig. 21.

While different methods for quantifying the damping have their specific theoretical basis, it remains unclear whether they inherently provide the same stability guarantees. Consequently, it is not clear whether passing one test of damping criterion naturally guarantees the pass of others, or stakeholders must go through all these different damping test procedures. This ambiguity creates additional complexity for both TSOs and OEMs, and thus, a more harmonized compliance criteria for damping assessment is urgently needed.

*2) Inertia*: It is important for TSOs to quantify the inertia contribution of GFM-ESSs. For SG, the inertia constant can be calculated based on the specific RoCoF and the corresponding active power, i.e.,

$$H = \frac{\Delta P}{2(d\omega/dt)}. \tag{3}$$

However, for GFM-ESSs, the active power response during a RoCoF event is influenced not only by virtual inertia control but also by droop control, damping control, and potentially other control loops. Consequently, the total active power output in response to a frequency disturbance is a composite of inertia power, droop-based power, damping power, and other control-inducted active power. Consequently, following the conventional inertia constant calculation method used for SGs cannot yield accurate results for GFM-ESSs, as reported by National Laboratory of the Rockies [28], see Fig. 22.

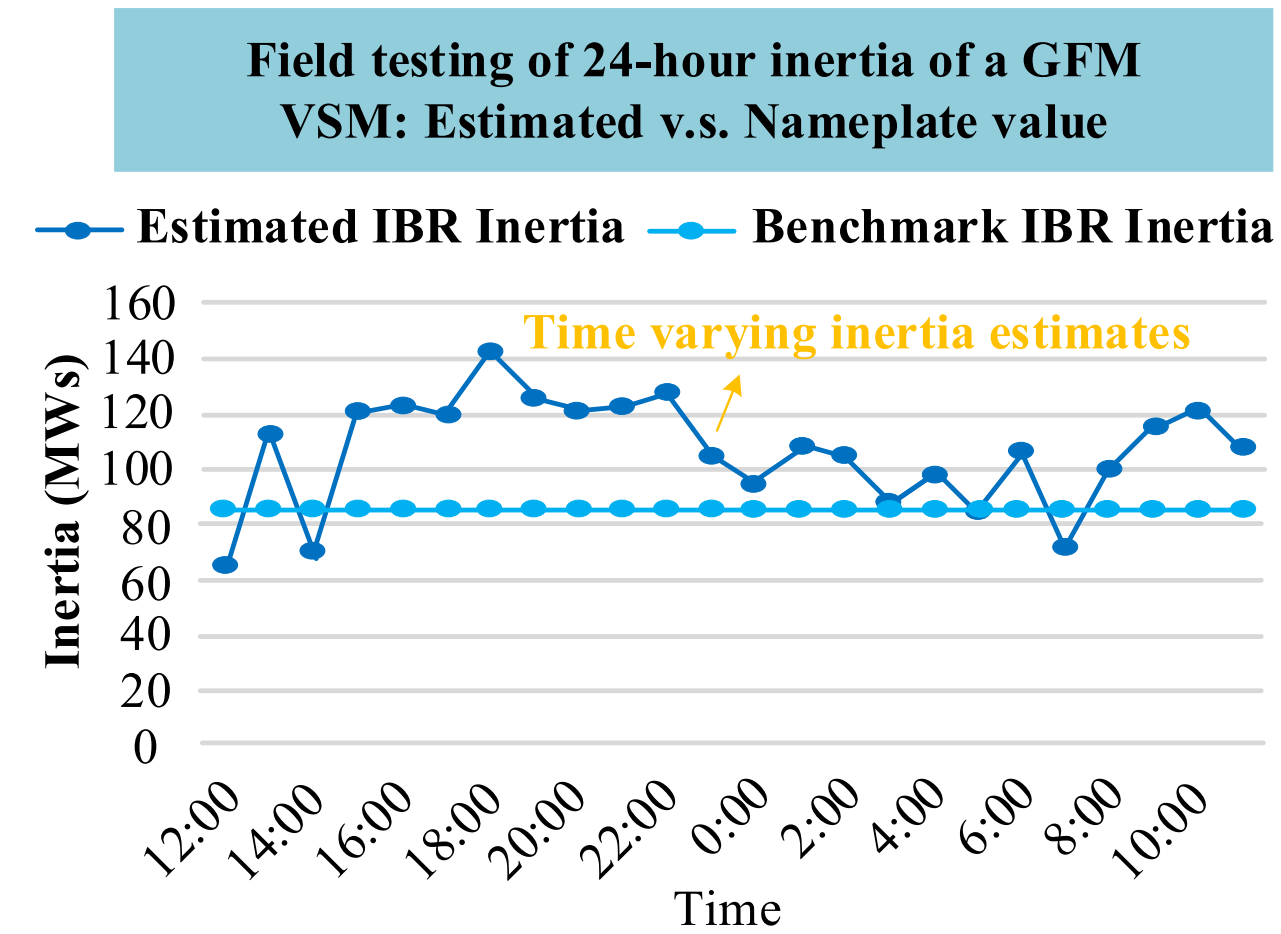

Fig. 22. Inaccurate inertia constant estimation, source:[28].

To address this challenge, it is suggested to deactivate all control functions except the virtual inertia control during inertia constant measurements [22]. However, this approach presents a significant drawback: disabling the damping control tends to destabilize the GFM-ESS. Therefore, OEM may reduce the gain of damping control, rather than fully disabling it during the inertia constant measurements, which, however, would still affect the measurement accuracy.

An alternative method for estimating the inertia constant is through the network frequency perturbation (NFP) plot [25]-[26]. The NFP plot is the bode diagram of the GFM converter's transfer function that characterizes its active power response under grid frequency perturbations, i.e.,

$$R_{NFP}(s) = \frac{\Delta P_o(s)}{\Delta f_g(s)}. \tag{4}$$

As pointed out by National Energy System Operator (NESO) [25], there is an approximate one-to-one mapping between the profile of NFP plot in different frequency ranges and the parameters of GFM-CBRs, e.g., inertia constant, damping, etc. This correlation enables the estimation of inertia constant based on the profile of the NFP plot. However, the interpretation guide of the NFP plot in [25] is based on the oversimplified small-signal model of GFM-CBRs, which is not always applicable in practice – particularly when vendor specific GFM control that might deviate from the conventional swing equations of SGs [29].l

*3) Performance under the limit*: While recent grid codes and technical specifications have clearly specified how current-limiting control should be implemented, the assessment of the dynamic performance of GFM-ESS when various limits are reached (current limit, energy limit, DC voltage limit, etc) remains a significant challenge for TSOs, as indicated by option F of Q8. Recently, there are several discussions on the inertia support capability of GFM-ESS considering current and power limit. For example, it is pointed out by the Australian Energy Market Operator (AEMO) that the inertial contribution from the GFM-ESS is no longer constant as that of SGs, but rather "*varies depending on various factors such as operating point, and overload capability. The inertia capability curve may be asymmetric depending on the operating point, magnitude and the direction of frequency event, available headroom and charging and discharging capacity.*" [30], as shown in Fig. 23. Similar insights have also been shared by VDE FNN, who has further differentiated the inertial contribution into "positive inertia (power increase during negative RoCoF)" and "negative inertia (power decrease during positive RoCoF)" depending on the direction of the frequency event [27].

However, how to evaluate the damping contribution of GFM-ESS under various limits remains an open issue. Unlike conventional SGs, where damping is relatively well-defined by machine parameters and excitation system characteristics, GFM-ESS exhibits nonlinear and state-dependent behavior when operating at its limit. As a result, TSOs lack practical tools and criteria to certify whether a GFM-ESS will still provide adequate damping during extreme but credible events.

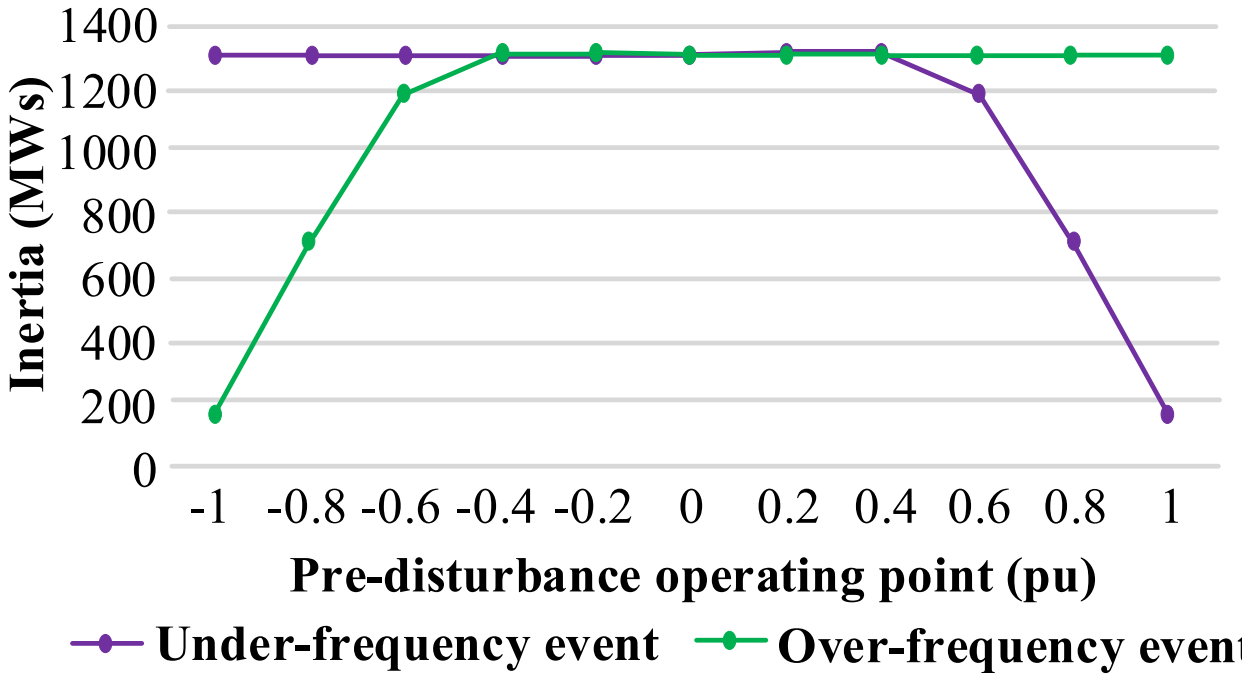

Fig. 23. Various inertia contribution of GFM-ESS depending on the direction of frequency event. Source: [30]

*B). Market*

It can be seen from the response of Q11 that option D "*New energy market with the consideration of ancillary services provided by GFM-ESS.*" is identified as the second most emerging task for accelerating the deployment of GFM-ESS. This is because GFM-ESS is generally more expensive than its GFL counterparts, and thus, a proper market design is very important to provide economic incentives for different stakeholders to deploy this technology. However, translating GFM capabilities into commercially viable, market-based stability services remains a significant challenge across multiple jurisdictions. In the following, the current attempts and remaining challenges are summarized.

*1) Current Attempts:*

*1.1) Germany:* Germany has already launched market-based procurement of inertia, in which GFM-CBR that fulfill VDE FNN technical requirements can participate [27]. The inertia products are categorized as "positive inertia (power increase during negative RoCoF)" and "negative inertia (power decrease during positive RoCoF)". Therefore, not only GFM-ESS, but also GFM-PV/wind can participate the inertia market (difficult to provide positive inertia due to the limited additional energy/power from the source side, but can easily provide negative inertia). Moreover, for both positive and negative inertia, two availability-based product tiers are defined: a "basic" product requiring at least 30% availability, and a "premium" product requiring at least 90% availability [31]. This tiered structure provides flexibility for different asset types and business models.

*1.2) United Kingdom:* The United Kingdom has pioneered the procurement of stability services through its Stability Pathfinder programme, a series of trial processes launched by NESO. The programme aimed to create competitive markets as part of the transition to a carbon-zero power system [32].

In Stability Phase 1, only synchronous condensers were contracted. However, Stability Phase 2 (Scotland) marked a significant milestone: contracts were awarded to both synchronous machines and GFM converter-based solutions, including 5 BESSs with GFM capabilities.

The market design features competitive tenders with contracts lasting 6 to 10 years to stimulate investment in new capacity. Services are defined as inertia, short circuit level (SCL), and dynamic voltage support, with remuneration based on availability payments (£/settlement period). Both synchronous machine and converter-based technologies (GFM-CBRs) are eligible [32].

Following the Pathfinder rounds, NESO progressed toward an enduring stability market. However, in stability Market Round 2, only synchronous condensers and open cycle gas turbines secured 7.3 GVA of inertia contracts. This shows that more work needs to be done in this area such that GFM will be part of the overall solutions to support system stability [33].

*1.3) Australia:* Australia is at the forefront of deploying GFM technology. However, it has adopted a cautious and staged approach to the associated market design. At present, minimum secure levels of inertia are determined as part of system security requirements and are delivered through regulated procurement mechanisms. These include transmission network investments and contracted services, with ancillary service frameworks such as Network Support and Control Ancillary Services (NSCAS) playing a supplementary role where required [34]. GFM based resources, including GFM-ESS, can participate in providing such services subject to technical assessment and approval processes.

Beyond the procurement of minimum inertia requirements, a dedicated spot market for inertia has not been introduced. This reflects an assessment that existing system security frameworks—particularly system strength investments, which provide inertia as a co-benefit—are expected to meet foreseeable inertia needs without introducing additional market complexity [35].

As a result, the introduction of a standalone inertia market has been deferred. This allows time for the current framework to be embedded and for operational experience with non-traditional sources of inertia, including GFM-CBRs, to mature before considering more advanced market-based arrangements.

*2). Challenges.*

It is evident that the market for GFM-CBRs is far from mature. One of the biggest challenges, as pointed out by Energinet [36], lies in translating GFM capabilities into market-based products. This is because a market service must be clearly defined, measurable, and verifiable – requirements that are particularly challenging for GFM-CBRs due to the following reasons:

1. Capability coupling: Multiple GFM services may be yielded under a single disturbance event. For example, during a RoCoF event, not only the active inertia power, but also active damping power and active droop-based power are simultaneously generated, as discussed in Section III-A. Consequently, developing separate market products (e.g., a dedicated inertia market) for each individual GFM service becomes challenging, since the responses cannot be easily decoupled.

2. Nonlinear behavior under limits: The response of a GFM-CBR becomes inherently nonlinear when it encounters operational limits – whether power, energy, DC-link voltage, or current limits. As a result, the actual delivered response depends strongly on remaining headroom, operating conditions, severity of the disturbances and even geographical location. This dependence complicates market design, which typically expects predictable and consistent performance across different scenarios.

To overcome these challenges, future efforts can focus on: (i) Developing standardized testing methodology that quantify or even isolate the coupled responses of GFM-CBR; (ii) Establishing dynamic performance metrics that account for nonlinear behavior of GFM-CBRs under the various limit, and (iii) Designing market frameworks that move beyond simple capacity-based products toward availability-based mechanisms that reflect the true system benefit of GFM resources.

### *C). Requirement of Overloading Capabilities*

Improving transient overloading capability of GFM-CBR could bring in many benefits, which are:

1) *Protection:* While the reliable operation of differential and distance relays, which are widely used in transmission systems, is not contingent upon a high fault-current magnitude, with fault-current phase angles and sequence components playing a more decisive role in protection performance [36]–[37], a sufficiently large fault current is still required to ensure the pickup and activation of these protection functions. Therefore, enhancing the transient overloading capability of GFM-CBRs remains relevant. This becomes even more important in distribution systems, where overcurrent relays are widely used and their reliable operation directly depends on sufficiently high fault currents.

2) *Less mode switching, better network support capability:* It can be seen from the response of Q8 that uncertain and degraded performance of GFM-CBR under current-limiting mode is among the most challenging issues for TSOs. With improved transient overloading capability, fewer GFM-CBRs switch to the current-limiting mode during grid faults, thereby maintaining the voltage-source behavior and providing more predictable responses. Additionally, the higher fault current injection from GFM-CBRs also implies the stronger support to the grid during severe disturbances.

Despite these benefits, improving the transient overloading capability of GFM-CBRs inevitably increases the overall system cost, creating a trade-off between technical and economic benefits. Consequently, whether overloading capability should be standardized remain controversial. As shown in the responses to Q12, 55% of respondents favor standardization to simplify design and analysis, while 40% argue that optimal overloading levels are project-specific and should not be mandated. In practice, different regions have adopted varying approaches: China has mandated a 3 pu/10 s overloading capability for GFM-BESS [39], whereas the AEMO defines overloading capability as a Type 2 transitional service to be procured through contracting with the service provider [40].

### *D). Accelerated Aging of Batteries*

The battery of a GFM-ESS may experience more frequent charging and discharging cycles due to the provision of ancillary services such as phase-jump power and inertia power. This increased cycling could potentially accelerate battery aging. However, as reflected in the responses to Q16, there is currently no consensus on the extent to which these ancillary services affect battery lifetime.

Specifically, approximately 30% of respondents selected option A, believing that phase-jump power and inertia services result in only minor charging and discharging cycles, making their impact on battery degradation negligible. They argue that the major degradation factors arise from other ancillary services that demand more power and energy, such as frequency restoration reserves (FRR) and frequency containment reserves (FCR). These services, however, can also be provided by GFL-ESS, suggesting that GFM control does not inherently influence battery aging beyond the additional phase-jump and inertia responses.

Conversely, 20% of respondents selected option B, stating that while phase-jump power and inertia services involve relatively small cycles in terms of time and energy, the transient overloading can significantly accelerate battery aging. This is because charging and discharging the battery at high C-rates substantially speeds up degradation [41].

Notably, about 37% of respondents chose option C, underscoring the need to quantify the occurrence rate and magnitude of both large- and small-signal events to accurately evaluate the influence of GFM operation on battery degradation. This perspective advocates an event-based assessment framework in place of blanket assumptions regarding battery aging.

Taken together, these responses indicate that while battery aging is a legitimate concern, its practical significance remains contested. Developing standardized event profiles and battery degradation models are therefore needed to accurately assess the lifecycle costs associated with GFM-ESSs and to inform the design of suitable market compensation frameworks.

### *E). Summary of the Challenges*

The identified challenges of GFM-ESS are summarized in Table VII.

TABLE VII
SUMMARIZE OF THE CHALLENGES

| | Inertia | Damping | Fault Current |
|---|---|---|---|
| **Feature** | • Tunable (can exceed SG) if limits* are not reached<br>• Reduced and nonlinear when reaching the limits | • Tunable (can exceed SG) if limits are not reached<br>• Reduced and nonlinear when reaching the limits | • Natural response ensures phase angle/sequence similar to SG at no extra cost<br>• Higher overloading brings in benefits but adds cost |
| **Quantification** | Lack of quantification method:<br>• RoCoF-based test couples inertia, damping, and droop responses. Disabling damping risks system stability [28].<br>• NFP plot-based inertia estimation is based on the oversimplified small-signal model, which is not always applicable [29] | Can be quantified but the criteria are not unified:<br>• Impedance-based damping assessment [24]<br>• Damping torque based assessment [25]-[26]<br>• Damping assessment based on time domain response [23],[27] | • Easily quantifiable via current measurement |
| **Market Design** | Capability coupling and nonlinear behavior under the limits complicate the market design for GFM services | | |
| **Battery Aging** | No consensus if GFM service would accelerate battery aging:<br>• Opinion A: Phase-jump and inertia cause minor cycles → negligible degradation<br>• Opinion B: Transient overloading (high C-rate) significantly accelerates aging [41] | | |

*Here the limits refer to energy/power/DC-link voltage/current limit

## V. CONCLUSION AND FUTURE ACTIONS

Through the global survey initiated by Cigre B4.101, state-of-the-art consensus and remaining challenges for deploying GFM-ESS are identified. The findings are anticipated to provide a valuable foundation for future research and the development of standardization frameworks. In the coming years, Cigre B4.101 will organize more technical activities to facilitate the collaborations between academia and industry in addressing the challenges, which can hopefully contribute to maturing the technology for stabilizing the future CBR-dominated power systems.

## VI. ACKNOWLEDGEMENT

The authors would like to express sincere gratitude to all expert members of Cigre Working Group B4.101 for their dedicated contributions to the design of the questionnaire used in this survey. The authors would like to also express the appreciation to all survey participants for taking the time to share their valuable insights, experiences, and perspectives. Special thanks go to Joanne Hu from RBJ Engineering Corporation, for her efforts in reviewing the paper and providing constructive comments.

## APPENDIX

The detailed statistics of the survey participants are given in Table A-I to Table A-VI.

TABLE A-I
RESPONDENT FROM ACADEMIA

| Institution |
|---|
| • Massachusetts Institute of Technology<br>• University of Toronto<br>• Imperial College London |

- University of Birmingham
- KTH Royal Institute of Technology
- Aalborg University
- Delft University of Technology
- KU Leuven
- University of Wollongong
- Zhejiang University
- Shandong University
- Northeast Electrical Power University
- Sakarya University of Applied Science
- Bayern University
- Indian Institute of Technology Jammu
- University of South Africa
- University of Chile
- Georgia Technical University

TABLE A-II
RESPONDENT FROM TSO/DSO

| Institution |
|---|
| • TenneT TSO GmbH<br>• TransnetBW GmbH<br>• Amrpion<br>• Energinet<br>• Red Electrica<br>• RTE<br>• National HVDC Centre<br>• Fingrid<br>• AEMO<br>• State Grid Gansu Electric Power Company<br>• Saudi Electricity Company<br>• Nampower, GCC Interconnection Authority<br>• National Transmission Company South Africa<br>• Tramsmission System Operator (Albanian)<br>• Electricity generating authority of Thailand |

TABLE A-III
RESPONDENT FROM OEM

| Institution |
|---|
| • ABB<br>• Hitachi Energy<br>• GE Vernova<br>• Siemens Energy<br>• Siemens Gamesa Renewable Energy<br>• SMA Solar Technology AG<br>• NR Electric<br>• Huawei<br>• Sungrow<br>• China Railway Rolling Stock Corporation (CRRC)<br>• Beijing Sifang Automation Company<br>• Hopewind<br>• NHOA Energy<br>• Fluence<br>• Engineering Specialist - Microgrid Control Platforms<br>• KONČAR |

TABLE A-IV
RESPONDENT FROM RESEARCH INSTITUTE

| Institution |
|---|
| • SuperGrid Institute<br>• L2EP - Centrale Lille<br>• ITE Instituto Tecnológico de la Energía<br>• China Electric Power Research Institute<br>• China Southern Power Grid Electric Power Research Institute<br>• State Grid Jiangsu Electric Power Company Research Institute<br>• State Grid Jibei Electric Power Research Institute<br>• Korea Electrotechnology Research Institute<br>• Central Research Institute of Electric Power Industry (Japan) |

TABLE A-V
RESPONDENT FROM DEVELOPER

| Institution |
|---|
| • Shell Global Solutions International<br>• RWE<br>• Equinor<br>• Zenobe<br>• Abraget<br>• Silicon Ranch |

TABLE A-VI
RESPONDENT FROM CONSULTANT

| Institution |
|---|
| • Manitoba Hydro International<br>• Aurecon,<br>• DNV<br>• Convergrid<br>• Motor Oil Hellas,<br>• Parvus<br>• Nyquist, Eng. and Consulting |